\documentclass[trackchanges, twocolumn]{aastex701}

\newcommand{\iasym}[2]{\ensuremath{^{+#1}_{-#2}}\xspace}

\usepackage{xspace} 
\usepackage{upgreek} 
\usepackage{booktabs}
\usepackage{threeparttable}
\usepackage{amsmath}
\usepackage{siunitx}

\begin{document}


\title{Asymmetric nightside CO$_2$ features, inefficient heat transport, and precise evolutionary constraints: Spectroscopic phase curves reveal the past and present of a white dwarf--brown dwarf binary 
}

\author[orcid=0009-0005-9531-7733, gname={Daphne}, sname={Broski-Laing}]{Daphne Broski-Laing}
\affiliation{Department of Astronomy, University of Virginia, Charlottesville, VA 22904, USA}
\email[show]{daphne.broski.laing@virginia.edu}

\author[orcid=0000-0003-2969-6040, gname={Yifan}, sname={Zhou}]{Yifan Zhou}
\affiliation{Department of Astronomy, University of Virginia, Charlottesville, VA 22904, USA}
\email[show]{yzhou@virginia.edu}

\author[orcid=0000-0003-3667-8633, gname={Joshua}, sname={Lothringer}]{Joshua D. Lothringer}
\affiliation{Space Telescope Science Institute, 3700 San Martin Drive, Baltimore, MD 21218, USA}
\email{jlothrin@stsci.edu}

\author[orcid=0000-0003-3714-5855, gname={Daniel}, sname={Apai}]{Daniel Apai}
\affiliation{Steward Observatory, University of Arizona, 933 N. Cherry Ave., Tucson, AZ 85721, USA}
\affiliation{Lunar and Planetary Laboratory, University of Arizona, 933 N. Cherry Ave., Tucson, AZ 85721, USA}
\affiliation{James C. Wyant College of Optical Sciences, University of Arizona, Tucson, AZ 85721, USA}
\affiliation{Earth, Atmospheric, and Planetary Sciences, Massachusetts Institute of Technology,  77 Massachusetts Avenue, Cambridge, MA 02139, USA}
\email{apai@arizona.edu}

\author[orcid=0000-0003-0825-4876, gname={Jenni}, sname={French}]{Jenni R. French}
\affiliation{School of Physics and Astronomy, University of Birmingham, Birmingham B15 2TT, UK}
\email{j.french.4@bham.ac.uk}

\author[orcid=0000-0003-2478-0120, gname={Sarah}, sname={Casewell}]{Sarah L. Casewell}
\affiliation{Centre for Exoplanet Research, School of Physics and Astronomy, University of Leicester, Leicester LE1 7RH, UK}
\email{slc25@leicester.ac.uk}

\author[orcid=0000-0002-4321-4581, gname={Laura}, sname={Mayorga}]{L. C. Mayorga}
\affiliation{Johns Hopkins University Applied Physics Laboratory, Laurel, MD 20723, USA}
\email{Laura.Mayorga@jhuapl.edu}

\author[orcid=0009-0003-6759-7067, gname={Lael}, sname={Shin}]{Lael Shin}
\affiliation{Steward Observatory and Department of Astronomy, University of Arizona, Tucson, AZ 85721, USA}
\email{lshin@arizona.edu}

\author[orcid=0000-0003-1487-6452, gname={Ben}, sname={Lew}]{Ben W. P. Lew}
\affiliation{Bay Area Environmental Research Institute, Moffett Field, CA 94035, USA}
\affiliation{NASA Ames Research Center, Moffett Field, CA 94035, USA}
\email{weipenglew@arizona.edu}

\author[orcid=0000-0003-2278-6932, gname={Xianyu}, sname={Tan}]{Xianyu Tan}
\affiliation{Tsung-Dao Lee Institute \& School of Physics and Astronomy, Shanghai Jiao Tong University, Shanghai 201210, China}
\email{xianyut@sjtu.edu.cn}

\author[orcid=0000-0001-9521-6258, gname={Vivien}, sname={Parmentier}]{Vivien Parmentier}
\affiliation{Laboratoire Lagrange, Observatoire de la C\^{o}te d'Azur, Universit\'{e} C\^{o}te d'Azur, Nice, France}
\email{vivien.parmentier@oca.eu}

\author[orcid=0000-0002-8808-4282, gname={Siyi}, sname={Xu}]{Siyi Xu}
\affiliation{NSF NOIRLab, 950 N.\ Cherry Avenue, Tucson, AZ 85719, USA}
\email{siyi.xu@noirlab.edu}

\author[orcid=0000-0002-5251-2943, gname={Mark}, sname={Marley}]{Mark S. Marley}
\affiliation{Lunar and Planetary Laboratory, University of Arizona, Tucson, AZ 85721, USA}
\email{marley@lpl.arizona.edu}


\begin{abstract}

We present the first \textit{JWST} phase curve of a white dwarf-brown dwarf binary, a NIRSpec PRISM observation of ZTFJ0038+2030. Short-period white dwarf--brown dwarf binaries provide unique laboratories to probe substellar atmospheres. Tidal locking drives hot Jupiter-like atmospheric dynamics in the brown dwarf. The system's formation history offers a window into planetary systems around post-main-sequence stars. We obtain a full-orbit phase curve of ZTF0038, including a total eclipse of the white dwarf, which enables us to separate the two components' emission throughout the entire orbit, and we model the brown dwarf's phase-resolved emission spectra using substellar atmosphere forward models and atmospheric retrievals. The PRISM spectrum covers $\sim$80\% of the brown dwarf's bolometric emission, enabling a nearly model-independent energy balance calculation, which yields a day-to-nightside heat transport efficiency of $\lesssim10\%$. Inefficient heat redistribution is further supported by the phase curve shape and the nightside spectrum closely resembling non-irradiated mid-to-late T dwarfs. The spectroscopic phase curves reveal a stark nightside asymmetry associated with a strong CO$_2$ absorption feature at 4.2 $\upmu$m, while the retrieved abundances indicate a longitudinally homogeneous distribution of CO$_2$ as well as all other key species detected in the atmosphere. The precise internal luminosity measurement of the brown dwarf informs both the age of the WD--BD system (7.5--8.8 Gyr) and indicates a low common-envelope ejection efficiency. These data illustrate the exquisite opportunity to probe the three-dimensional processes of substellar atmospheres, connect substellar and exoplanet atmospheres, and probe the evolution of post-main-sequence planetary systems using WD--BDs.

\end{abstract}


\section{Introduction}\label{sec:intro}

White dwarf--brown dwarf (WD--BD) binaries are rare survivors of common envelope evolution that carry the imprints of close binary interaction and the subsequent evolution of irradiated substellar companions \citep[e.g.,][]{Maxted2006Natur.442..543M, Farihi2005AJ....130.2237F, Steele2007MNRAS.382.1804S, Casewell2012ApJ...759L..34C, Casewell2015MNRAS.447.3218C,Walters2023}. These systems form when a brown dwarf in the brown dwarf desert becomes engulfed by its host star's expanding envelope during post-main-sequence evolution. Drag during the common envelope phase causes orbital decay, and the orbital energy transferred to the envelope drives its ejection \citep{Zorotovic2017ApJ...846..117Z, zorotovic_close_2022}. The brown dwarf is left tidally locked in an ultra-short period orbit, with known systems spanning periods from 70 minutes to a little over 10 hours \citep{Steele2007MNRAS.382.1804S, 
Casewell2015MNRAS.447.3218C, zhou_hstwfc3_2021}. The resulting permanent day-night temperature contrast drives the same atmospheric processes observed in hot Jupiters — circulation, vertical mixing, chemical quenching, cloud formation, and photochemical reactions \citep{Lothringer2020, Showman2020, TanShowman2020ApJ...902...27T, Lee2020}. These binaries are unique laboratories for testing atmospheric theories in physically distinct circulation regimes. Their irradiation, rotation rates, and interior heat flux exceed those of hot Jupiters by orders of magnitude \citep{TanShowman2020ApJ...902...27T, Lee2020}.

Phase curves offer the most direct observational probe of the three-dimensional structure of tidally locked atmospheres \citep{Knutson2007, Parmentier2018haex.bookE.116P, Stevenson2014}. As different hemispheres rotate into view, phase curves constrain the longitudinal distribution of temperature, clouds, and chemical composition \citep{Parmentier2018haex.bookE.116P}. Observations with \textit{Spitzer} and \textit{Kepler} established the basic framework: hot Jupiters generally show large day-night flux contrasts and an eastward offset of the brightness maximum, consistent with equatorial superrotation \citep{Knutson2007, Showman2009ApJ...699..564S}. Spectroscopic phase curves further enhance the constraints on circulation, disequilibrium chemistry, cloud opacity, and photochemistry \citep[e.g.,][]{Stevenson2014}. By mapping the phase-resolved emission spectrum across a broad wavelength range, they simultaneously constrain the day-night temperature contrast, the longitudinal distribution of molecular abundances, cloud opacity, and the signatures of photochemical products \citep{Stevenson2014, zhou_hstwfc3_2021, Amaro2025}.

White dwarf--brown dwarf binaries are particularly powerful targets for spectroscopic phase curve observations \citep{Casewell2015MNRAS.447.3218C, zhou_hstwfc3_2021, Lothringer2024}. Their ultra-short orbital periods of 1 to 10 hours enable complete phase coverage in a single observing sequence. The extreme irradiation produces day-night flux modulations of 10\%–50\% in the near-infrared, more than an order of magnitude larger than typical hot Jupiter phase curve amplitudes \citep{zhou_hstwfc3_2021, Amaro2025, Lew2022}. 
Because the brown dwarf radius is typically $\sim10$ times larger than their white dwarf hosts, the flux ratios are highly favorable, nearing and sometimes exceeding 1 in the infrared. The white dwarf's spectrum is nearly featureless, typically non-variable (an exception is \cite{Amaro2025}), and fades at longer wavelengths, facilitating clean component separation. \textit{Spitzer} and \textit{HST}/WFC3 phase curve observations of multiple WD--BD systems have confirmed these advantages and provided initial constraints on day-night energy transport \citep{zhou_hstwfc3_2021, Amaro2025, Lew2022, French2024}. However, WFC3's wavelength coverage from 1.1 to 1.7 $\upmu$m
 probes only a single water band and encloses less than 30\% of the total thermal emission and only covers one major molecular absorption feature (H$_2$O at 1.4 $\upmu$m). Furthermore, the anomalous flux excesses observed at K-band and Spitzer 4.5\,$\upmu$m in some brown dwarfs remain unexplained, pointing to unknown circulation or photochemical effects that demand broader spectral coverage to resolve \citep{Casewell2015MNRAS.447.3218C, Lew2022}.

In JWST Cycle 3 (GO-4967, \citealt{Zhou2024jwst.prop.4967Z}), we are conducting NIRSpec PRISM bright object time-series observations of five WD--BD binaries spanning equilibrium temperatures from ${\sim}$750 K to ${\sim}$3300 K. This temperature range straddles the warm to ultra-hot Jupiter regimes. The 0.6--5.3 $\upmu$m wavelength coverage captures  $\sim$80\% of the thermal emission from each brown dwarf and simultaneously resolves molecular features from H$_2$O, CH$_4$, CO, CO$_2$, and SO$_2$.  This program constrains how competing external irradiation and internal heat flux govern atmospheric structure, how common envelope evolution alters companion properties, and how these systems compare to isolated brown dwarfs and hot Jupiters across multiple physical parameters. ZTF0038B \citep{van_roestel_ztfj00382030_2021}, the coolest brown dwarf in our sample and among the coolest known in any WD--BD system, uniquely bridges the atmospheric regimes of hot Jupiters, isolated brown dwarfs, and solar system gas giants.

\object[GALEX J003854.9+203026]{ZTF J0038+2030AB} (hereafter ZTF0038) is an eclipsing binary with a 10.4 hour orbital period hosting the coolest brown dwarf companion in any known WD--BD system \citep{van_roestel_ztfj00382030_2021}. The host is a 10,900 K, 0.505 M$_\odot$ white dwarf, and the brown dwarf is a transiting 62 M$_\text{Jup}$ companion with no previous IR emission detections before these \textit{JWST} observations. The white dwarf has a radius of 0.01429 $R_\odot$ and a surface gravity of $\log g = 7.832$. The brown dwarf has a radius of 0.0783 $R_\odot$, a surface gravity of $\log g = 5.425$, and an equilibrium temperature of $T_\text{eq} = 745$ K. The connection across regimes is not merely qualitative — each physical parameter places ZTF0038B squarely within a distinct class: its rotation period and external-to-internal heat flux ratio resemble solar system giants, its bolometric irradiation flux is comparable to warm Jupiters, and its interior heat flux places it among L/T transition brown dwarfs. This paper presents the first \textit{JWST} NIRSpec PRISM spectroscopic phase curve of ZTF0038B, which includes a total eclipse of the white dwarf by the brown dwarf.

This paper is organized as follows. Section \ref{sec:obs} describes the observations and data reduction. Sections \ref{sec:lc_analysis} and \ref{sec:spectroscopic_analysis} present the light curve and spectroscopic analyses. Section \ref{sec:CO2_asymm} presents the anomalous light curve behavior observed at the 4.2 $\upmu$m CO$_2$ absorption feature. Section \ref{sec:energy_balance} presents the energy balance calculation. Section \ref{sec:discussion} discusses circulation and evolutionary properties of the brown dwarf, and Section \ref{sec:conclusion} summarizes our conclusions.

\section{Observations and Data Reduction} 
\label{sec:obs}
\subsection{JWST Time-Series Observations}
ZTF0038 is the first target observed in a NIRSpec PRISM phase curve survey of five WD--BD binaries (GO-4967, PI: Yifan Zhou). The 10.6 hour observation was taken between UTC 2024-12-04 04:04:58 and UTC 2024-12-04 14:41:37.623. The observation covers 102\% of the 10.4 hour orbital period \citep{van_roestel_ztfj00382030_2021}.

The near-IR spectrum of ZTF0038 was monitored using the NIRSpec PRISM bright object time-series mode (BOTS). The observations were split into four exposures with 208 integrations per exposure, and 50 groups per integration. The integration time is 45.232\,s. The spectrum continuously covers 0.6 to 5.3\,$\upmu$m with a spectral resolution $R\sim100$ at 2.95 $\upmu$m. 

The observed flux decreased rapidly as the brown dwarf eclipsed the white dwarf, and we observed the total eclipse for 8 integrations while the brown dwarf fully covered the white dwarf. Three integrations before and after totality captured ingress and egress. During totality, we observed the brown dwarf's nightside spectrum without white dwarf contamination.

\subsection{Time-Series Spectra Extraction}

\begin{figure*}
     \centering
     \includegraphics[width=0.8\linewidth]{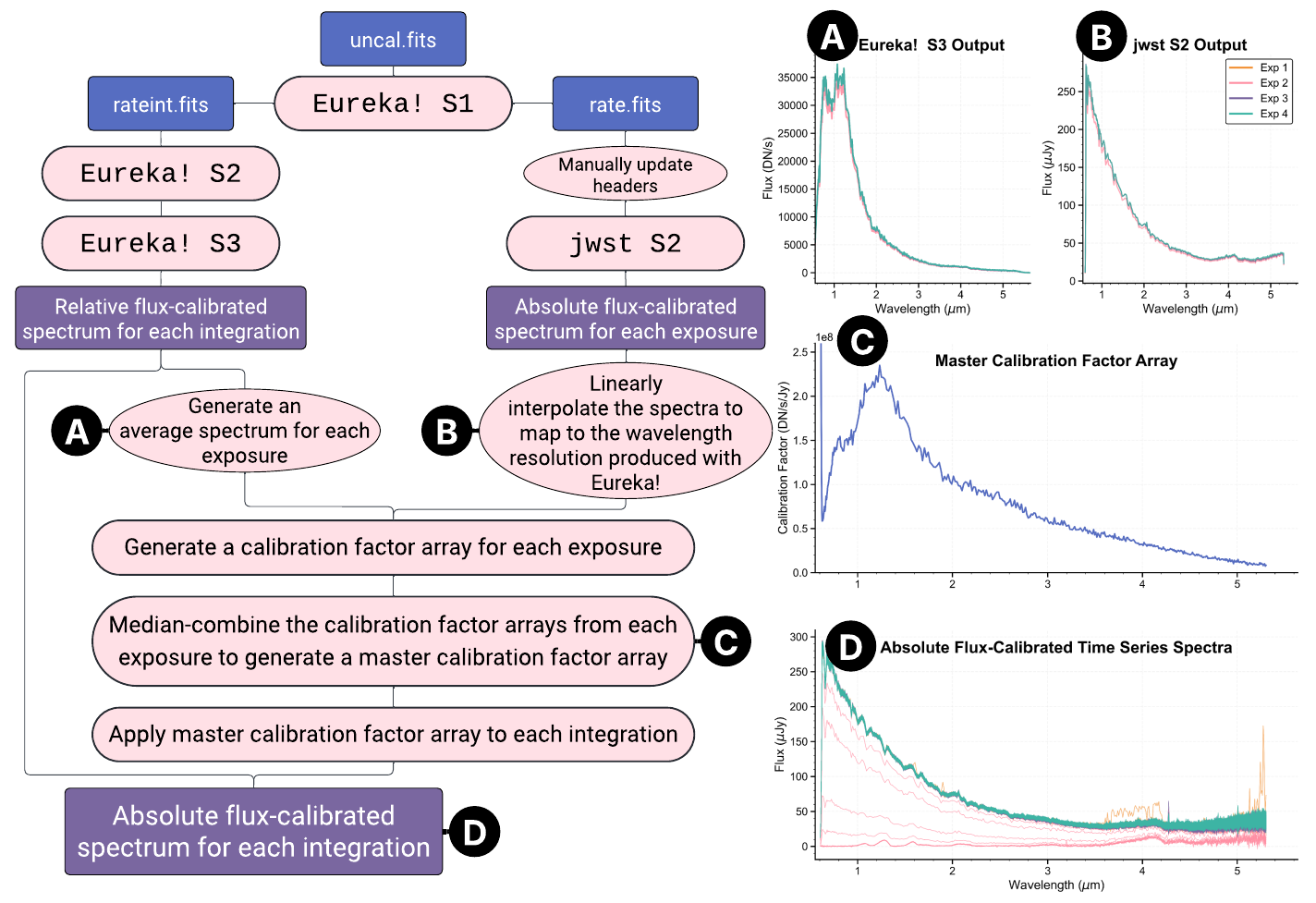}
     \caption{\emph{Left:} Outline of the custom spectral extraction pipeline used to generate absolute flux-calibrated time-series spectra. \emph{Right:} intermediate data products generated at various stages throughout the reduction process: (A) relative flux-calibrated spectral time series produced by \texttt{Eureka!}; (B) flux-calibrated and exposure-averaged spectra produced by the \texttt{jwst} pipeline; (C) master calibration factor array; and (D) absolute flux-calibrated spectral time series. These time-series spectra include a few visible outlier integrations which are noted in Table \ref{tab:eclipse_frames} and excluded from all subsequent analysis.}
     \label{fig:pipeline_flowchart}
 \end{figure*}

To optimize the extraction of flux-calibrated time-series spectra, we developed a custom spectral extraction pipeline that uses two data processing packages: \texttt{jwst} \citep{Bushouse2023} and \texttt{Eureka!} \citep{Bell2022}. The \texttt{jwst} Python package is the standard data reduction pipeline for \textit{JWST} observations, and \texttt{Eureka!} is well-vetted by the transiting exoplanet community for optimizing space-based relative photometry and light curve measurements \citep{Feinstein2023,Carter2024}. We use the \texttt{jwst} package to determine the absolute flux density and the \texttt{Eureka!} package to preserve the precision of the spectroscopic variability. A set of time-independent calibration factors is derived to scale the \texttt{Eureka!} relative photometry to absolute flux units. Figure~\ref{fig:pipeline_flowchart} outlines our optimized method for extracting precise, absolute flux-calibrated time-series spectra. The pipeline is available on GitHub.\footnote{\url{https://github.com/djblaing/DBL_time-series_spectral_extraction}} The \textit{JWST} Science Data Processing version used to produce the uncalibrated 2D data files (\texttt{uncal.fits} files) was 2025\_1, and the version of the \textit{JWST} calibration pipeline used for all subsequent processing is version 1.16.1.

We use Stage 1 of \texttt{Eureka!} to generate \texttt{rate.fits} and \texttt{rateints.fits} files from the uncalibrated data (\texttt{uncal.fits}) downloaded from the Barbara A. Mikulski Archive for Space Telescopes \footnote{\url{https://archive.stsci.edu/}}. Stage 1 of \texttt{Eureka!} is a wrapper of Stage 1 of the \texttt{jwst} pipeline to convert group-level readout into count rate per integration images (\texttt{rateint.fits} images) as well as count rate per exposure images (\texttt{rate.fits} images). After generating 2D spectral images in Stage 1, we manually identify and mask spurious pixels that are missed by the pipeline by identifying outliers in the spatial and temporal dimensions. The \texttt{rateint.fits} files include the time-resolved information important for the analysis of ZTF0038. The \texttt{rate.fits} files are required to generate calibration factors using \texttt{jwst}.

The \texttt{rateint.fits} images are further calibrated and converted to 1D time-series spectra using \texttt{Eureka!} Stage 2. Stage 3 of \texttt{Eureka!} extracts the optimal time-series spectra and provides uncertainties propagated from the raw images. The resulting spectra from \texttt{Eureka!} Stage 3 provide the calibrated spectrum in units of DN/s. These spectra are not absolute-flux calibrated, but provide accurate relative flux changes between each spectrum throughout the observation.

   We use the \texttt{jwst} pipeline produce a time-averaged, absolute flux-calibrated spectrum. Some data reductions steps are automatically skipped for time-series observations\footnote{See the \texttt{jwst} documentation: \url{https://jwst-pipeline.readthedocs.io/en/stable/jwst/pipeline/calwebb_spec2.html}}. We bypass these hard-coded restrictions and avoid the additional complexities associated with time-series data processing by manually removing the time-series indicators. We produce a time-averaged \texttt{rate.fits} file for each exposure, and manually update the exposure type in the fits header from \texttt{NRS\_BRIGHTOBJ} to \texttt{NRS\_FIXEDSLIT}, a matching exposure type that is not associated with time-series observations. After these adjustments, Stage 2 of \texttt{jwst} produces four exposure-averaged absolute flux-calibrated spectra, one for each of the four exposures of ZTF0038.

We use these exposure-averaged spectra to generate a set of time-independent calibration factor arrays that are applied uniformly across every integration to convert the spectra produced by \texttt{Eureka!} to physical flux units. We generate four exposure-averaged spectra from the time-resolved spectra produced with \texttt{Eureka!}. We linearly interpolate the wavelength grid of the \texttt{jwst}-produced spectra to align it with the wavelength grid of the spectra produced using \texttt{Eureka!}. 

An array of calibration factors ($C_{\nu,\text{exp}}$) for each exposure is calculated as the relative flux-calibrated spectrum ($F_{\nu,\text{ exposure (\texttt{Eureka!})}}$) divided by the absolute flux-calibrated spectrum ($F_{\nu,\text{ abs cal exposure (\texttt{jwst})}}$):
\begin{equation} \label{eqn:cal_factor}
C_{\nu,\text{ exp}} = \frac{F_{\nu,\text{ exposure (\texttt{Eureka!})}}}{F_{\nu,\text{ abs cal exposure (\texttt{jwst})}}}.
\end{equation} All four calibration factor arrays are averaged to generate the master calibration factor array ($C_{\nu,\text{master}}$):
\begin{equation} \label{eqn:master_cal_factor}
C_{\nu,\text{ master}} = \langle C_{\nu,\text{ exp}} \rangle.
\end{equation}
The finalized, optimal flux-calibrated time-series spectra ($F_{\nu,\text{opt int}}$) are calculated by applying the calibration factor to the time-resolved spectra produced with \texttt{Eureka!}:
\begin{equation} \label{eqn:cal_optspec}
F_{\nu,\text{ abs cal integration}} = \frac{F_{\nu,\text{  integration (\texttt{Eureka!})}}}{C_{\nu,\text{master}}}.
\end{equation}

\begin{figure*}[th]
     \centering
     \includegraphics[width=1\linewidth]{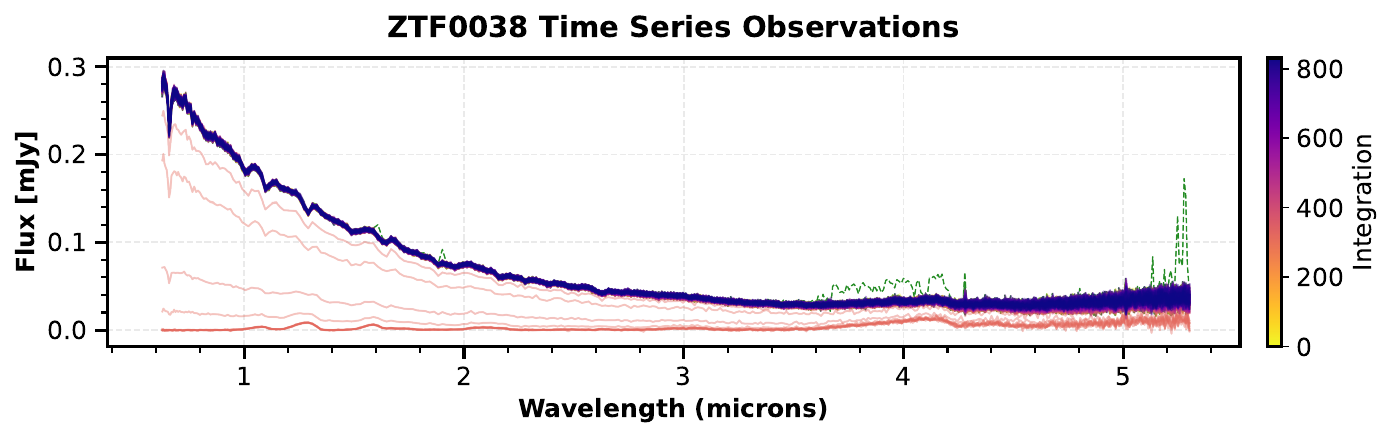}
     \caption{Absolute flux-calibrated spectra of the combined white dwarf--brown dwarf signal. As the brown dwarf eclipses the white dwarf, the flux decreases dramatically in the shorter wavelengths where the white dwarf signal dominates. The outlier integrations are shown in dashed green lines.}
     \label{fig:WDBD_spectra}
 \end{figure*}

\texttt{Eureka!} produces uncertainty estimates using the standard \texttt{jwst} data processing procedure. The calibration factor array, $C_{\nu,\text{master}}$, is the mean of the four exposures (832 integrations total). We exclude the uncertainty of the master calibration factor from our final spectral extraction because uncertainty is negligible compared to the per-integration uncertainty from \texttt{Eureka!}. At wavelengths between 1--5.3 $\upmu$m where the brown dwarf emits strongly, the relative uncertainty of the master calibration factor amounts to $\sim5\%$ of the per-integration uncertainty. This has a negligible effect on our analysis, and it is highly correlated with the individual uncertainties of each integration. The uncertainty of the final, absolute flux-calibrated spectra therefore scales by the master calibration factor $C_{\nu,\text{master}}$ (Equation \ref{eqn:cal_opterr}):
\begin{equation} \label{eqn:cal_opterr}
\sigma_{\nu,\text{ abs cal integration}} = \frac{\sigma_{\nu,\text{ integration (\texttt{Eureka!})}}}{C_{\nu,\text{ master}}}.
\end{equation}
The resulting spectra are shown in Figure~\ref{fig:WDBD_spectra}.

\begin{deluxetable}{lll}
\tablecaption{Integrations used in the data reduction and analysis of ZTF0038. WD + Nightside BD refers to integrations directly before and after the total eclipse, shown in Figure \ref{fig:isolating_BD}. The brown dwarf phase spectra are the 8 representative phases used for modeling and analysis.\label{tab:eclipse_frames}}
\tablecolumns{3}
\tablewidth{0pt}
\tablehead{
  \colhead{Label} & \colhead{Integrations} & \colhead{}
}
\startdata
Total Eclipse        & 315--322                    & \\
Partial Eclipse      & 312--314, 323--325          & \\
Systematic Outliers  & 76, 397, 495, 511, 705      & \\
WD + Nightside BD    & 308--311, 326--329          & \\
\hline
\multicolumn{2}{l}{\textbf{Brown Dwarf Phase Spectra}} & Phase\\
\hline
Nightside (A)          & 315--322           &  0   \\
Eastern Nightside (B)  & 416--422           & 0.125  \\
Eastern Terminator (C) & 518--525           & 0.25  \\
Eastern Dayside (D)    & 620--625           & 0.375\\
Dayside (E)            & 720--727           & 0.5 \\
Western Dayside (F)    & 10--17, 823--830   & 0.625 \\
Western Terminator (G) & 113--120           & 0.75 \\
Western Nightside (H)  & 213--220           & 0.875 \\
\enddata
\end{deluxetable}

The brown dwarf in ZTF0038 fully eclipses the white dwarf for eight integrations. During the three integrations prior to and following the eclipse, the flux of the brown dwarf changes rapidly within each 45 second integration, which interrupts the up-ramp-fitting procedures. We exclude the spectra observed during the partial eclipse from all analysis except the eclipse profile fitting. Through visual inspection, we identified five integrations that contain outliers, which are likely due to instrument systematics. These frames do not sample critical orbital phases and do not impact the analysis. We discarded these frames. The systematic outlier integrations as well as the integrations observed during the eclipse are presented in Table \ref{tab:eclipse_frames}.

\subsection{High-Cadence Spectral Extraction}\label{subsec:high_cadence_extraction}

Characterizing the eclipse profile requires increased temporal sampling to measure the rapid flux changes at ingress and egress. This is accomplished by using the group-level data available in the uncalibrated frames. Each 50-group integration is divided into five 10-group sub-integrations. This calculation is implemented by specifying the beginning and final group numbers for up-the-ramp fitting in the \texttt{jwst} Stage 1 pipeline. This procedure increases the temporal sampling rate by a factor of five. The high cadence data are reduced following the same procedure described in the spectral extraction step. The resulting broadband light curve is used to characterize the eclipse profile. The high-cadence data product is only used in eclipse profile fitting. The native cadence data was used in all other analyses.

\subsection{Separating the White Dwarf and Brown Dwarf Spectra}\label{subsec:separating_WD_and_BD}

When the brown dwarf eclipses the white dwarf, we observe the nightside spectrum of the brown dwarf without any contamination from the white dwarf. These observations correspond to the integrations indicated in Table \ref{tab:eclipse_frames}. We use these total eclipse observations as a tool to empirically separate the brown dwarf and white dwarf spectra across all phases. This procedure assumes the white dwarf is not variable, and our time-series observations validate this assumption. In the short wavelengths where the white dwarf emission dominates, the observations are consistent with a straight line model (see Section \ref{subsec:spec_pc_analysis}). 

We begin by median-combining all eight spectra extracted from the total eclipse integrations into a median-combined nightside spectrum for ZTF0038. We calculate empirical uncertainties for all median-combined spectra using Equation~\ref{eqn:median_combined_err}, where $\text{s}_{\nu,\text{combined}}$ is the standard deviation of the flux at each wavelength bin and $N_{\text{combined ints}}$ is the number of median-combined integrations.

\begin{equation} \label{eqn:median_combined_err}
\sigma_{\nu,\text{med spec}} = {\text{s}_{\nu,\text{combined}}}/{\sqrt{N_{\text{combined ints}}-1}}
\end{equation}

\begin{figure*}
    \centering
    \includegraphics[width=1\linewidth]{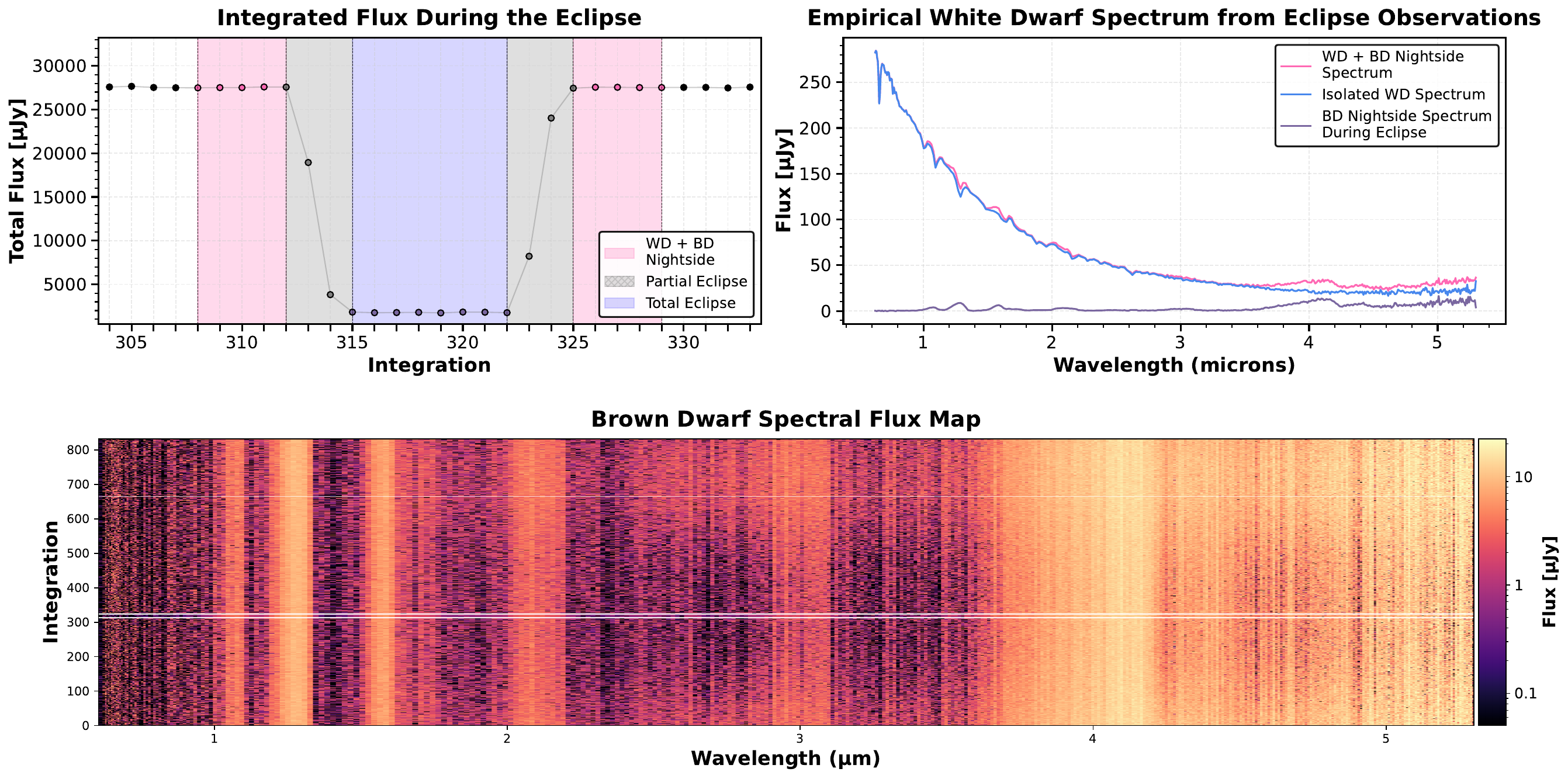}
    \caption{Overview of the procedure to separate the white dwarf and brown dwarf emission signals. The upper-left panel shows the broadband flux around the time of the eclipse. During the total eclipse (purple), we observe the brown dwarf's nightside emission without white dwarf contamination. Frames immediately before and after the eclipse (pink) contain combined signals from both objects. Subtracting the isolated nightside spectrum from these combined signals produces an empirical white dwarf spectrum (upper-right panel). Subtracting this white dwarf spectrum from all out-of-eclipse integrations yields isolated brown dwarf spectra for all of the integrations. The lower panel presents a 2D time-series spectrum. This spectral flux map presents wavelength on the x axis, integration number on the y axis, and spectral flux in the color map. The eight integrations during the ingress and egress are excluded from the dataset, and are shown as white horizontal lines. For the five systematic outlier integrations, we fill in the data by linearly interpolating between the neighboring integrations.}
    \label{fig:isolating_BD}
\end{figure*}

Next, we select a subset of out-of-eclipse integrations which contain the combined signal of the white dwarf and brown dwarf nightside observed immediately before and after the eclipse (Figure~\ref{fig:isolating_BD} and Table~\ref{tab:eclipse_frames}). These 8 integrations on either side of the eclipse span 16.7 minutes, corresponding to $\sim9.72^\circ$ of rotation. The brown dwarf's variability from the beginning of the selected out-of-eclipse frames until eclipse center is less than one sixth of the typical flux uncertainty of each data point. The brown dwarf's rotation therefore has a negligible impact on the observed emission for this interval. These out-of-eclipse integrations are median-combined with uncertainties calculated using Equation~\ref{eqn:median_combined_err}.

We subtract the brown dwarf's nightside emission observed during the eclipse ($F_{\nu,\text{total eclipse}}$) from the out-of-eclipse spectrum ($F_{\nu,\text{WD+BD, nightside}}$) to produce an isolated white dwarf spectrum ($F_{\nu,\text{WD}}$):

\begin{equation} \label{eqn:FWD}
F_{\nu,\text{ WD}} = F_{\nu,\text{ WD+BD nightside}} - F_{\nu\text{ total eclipse}}.
\end{equation}

The uncertainties of $F_{\nu,\text{WD}}$ are calculated by adding the uncertainties of $F_{\nu,\text{total eclipse}}$ and $F_{\nu,\text{WD+BD nightside}}$ in quadrature. Because the white dwarf shows negligible variability over the entire observation, we subtract $F_{\nu,\text{WD}}$ from all out-of-eclipse integrations to produce isolated brown dwarf spectra at every phase (Figure~\ref{fig:isolating_BD}, bottom panel). The uncertainties of the isolated brown dwarf spectra are calculated by adding the uncertainties of the out-of-eclipse spectrum and the white dwarf spectrum in quadrature.

\section{Light Curve Analysis}\label{sec:lc_analysis}

\subsection{Light Curve Production}\label{subsec:lc_production}

\begin{figure*}
    \centering
    \includegraphics[width=0.9\linewidth]{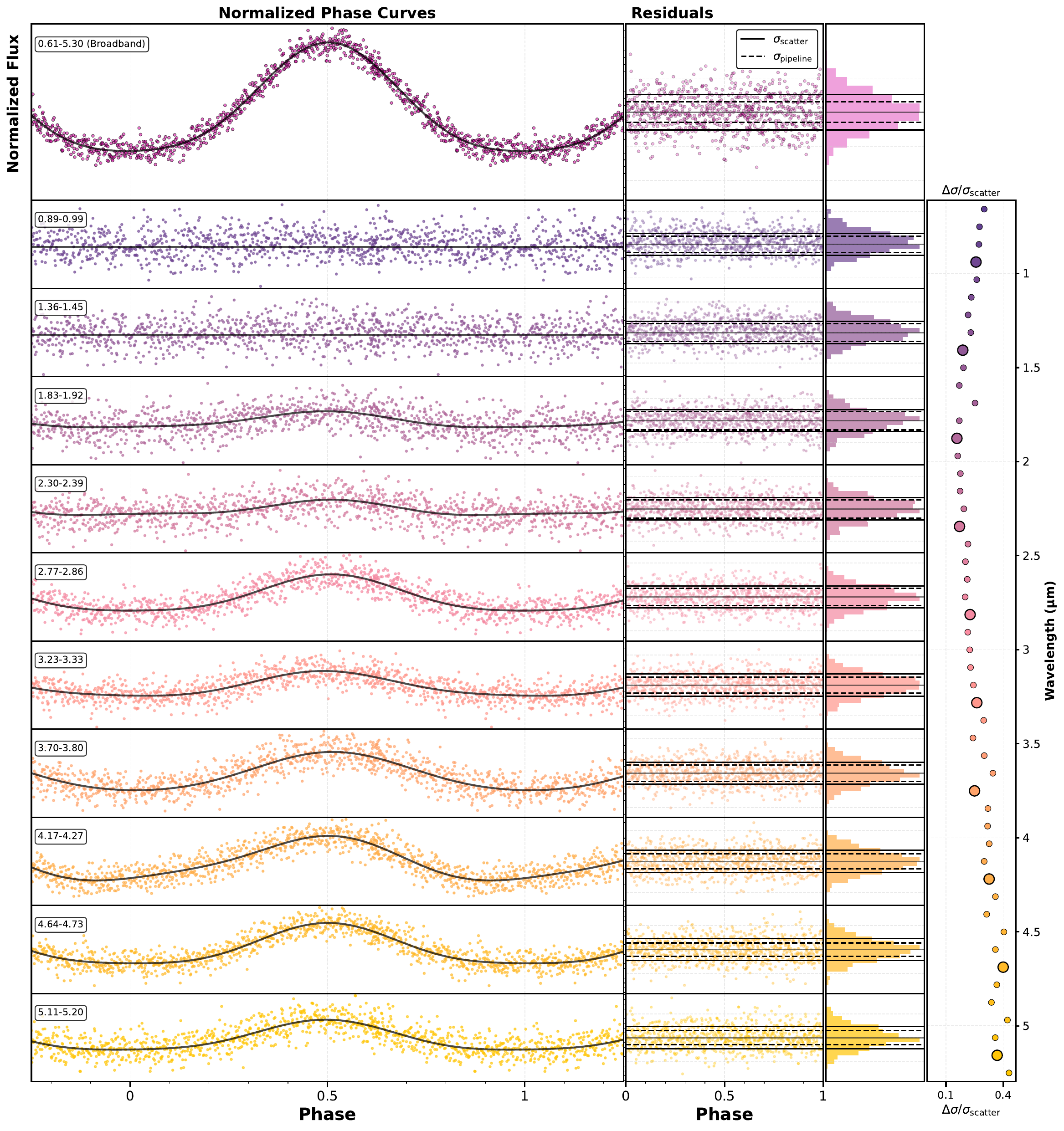}
   \caption{
   The broadband phase curve and a selection of 10 spectroscopic phase curves were produced for ZTF0038B (every 5 of the 50 evenly-spaced wavelength bins). Each phase curve is fit with a second order Fourier series with the period found from eclipse timing analysis (Table \ref{tab:binary_params_tab}). The residual columns show the phase curve residuals as a function of phase in the center column, and a histogram of the residuals to the right. Both residual panels overlay the pipeline uncertainty ($\sigma_\text{pipeline}$, dashed) and observed scatter ($\sigma_\text{scatter}$, solid), with $\Delta\sigma/\sigma_\text{scatter}$ quantifying the fractional excess scatter that is not accounted for by the pipeline. The right-most panel indicates the wavelength dependence of $\Delta\sigma/\sigma_\text{scatter}$, and the larger markers indicate the selection of 10 spectroscopic with phase curves shown on the left.}
    \label{fig:spec_pcs}
\end{figure*}

After subtracting out the white dwarf's contribution from the observed flux, we use the isolated brown dwarf spectra to generate broadband and spectroscopic phase curves for ZTF0038B. Each light curve is generated as the sum of $F_\nu$ across all wavelengths. The eclipse profile analysis utilizes the high-cadence data to generate a broadband light curve covering the $\sim$17 minutes surrounding the total eclipse of the white dwarf, while the broadband and spectroscopic phase curves utilize the full observation period. The broadband phase curve includes the summed flux from $0.6-5.3\,\upmu$m. For spectroscopic phase curves, we divide this wavelength range into 50 evenly-spaced bins. The broadband phase curve and a selection of the 50 spectroscopic phase curves are shown in Figure \ref{fig:spec_pcs}.

The pixel-level uncertainties are combined in quadrature to determine the uncertainty of the summed light curves. We assess these pipeline-produced uncertainties by comparing them with the observed scatter of the residuals in Figure \ref{fig:spec_pcs} (see Section \ref{subsec:spec_pc_analysis} for discussion of the phase curve models). The residuals are featureless, and therefore a reasonable and empirical estimate of the photometric uncertainty. We compare the standard deviation of the residuals ($\sigma_\text{scatter}$) and the RMS uncertainty of the isolated brown dwarf spectrum using the pipeline-produced uncertainty ($\sigma_\text{pipeline}$). We define $\Delta\sigma/\sigma_\text{scatter}$ to be the fraction of the observed scatter that is not accounted for by the pipeline: $\Delta\sigma/\sigma_\text{scatter} = (\sigma_\text{scatter} - \sigma_\text{pipeline}) /\sigma_{scatter}$. Each of these metrics is included in the supplementary data behind Figure \ref{fig:amplitude_phase_vs_wavelength}. We find that around 1/3 of the scatter is not accounted for by the pipeline, but  $\Delta\sigma/\sigma_\text{scatter}$ is wavelength-dependent, as shown in the right panel of Figure \ref{fig:spec_pcs}. This indicates that there are likely some mildly wavelength-dependent systematics that are not accounted for by pipeline-produced uncertainties.

Our broadband analysis uses the pipeline-produced uncertainties to determine a suitable phase curve model for the observations. Neither the absolute scaling of the uncertainty nor the mild wavelength dependent systematics impact our conclusion, so the pipeline-produced uncertainties are sufficient. However, our eclipse profile fitting and spectroscopic phase curve analysis uses the light curve uncertainties to produce uncertainties on fit parameters, and this requires a more thoughtful assessment of the time-series uncertainties. 

In order to accurately reflect the uncertainty of these light curves, we use the preferred model presented in Section \ref{subsec:broadband_pc_analysis}. For each spectroscopic phase curve, we re-scale the original pipeline-produced uncertainties with an error inflation factor,  $f_\text{inflate}$. The inflation factor is calculated as:

\begin{equation}
    f_\text{inflate} = \sqrt{\chi^2_{\text{r, pipeline }\sigma}}
\end{equation}

where $\chi^2_\text{r, pipeline}$ is the reduced chi-squared goodness-of-fit statistic calculated using the pipeline-produced uncertainty. By adopting this inflation factor, the reduced chi-squared statistic generated for the inflated uncertainties ($\chi^2_{\text{r, inflated }\sigma}$) becomes 1 by definition.

\begin{equation}
      \sigma_{\nu\text{, inflated}} = f_\text{inflate} \cdot \sigma_{\nu\text{, pipeline}}
\end{equation}

\begin{equation}
      \chi^2_{\text{r, inflated }\sigma} \equiv 1
\end{equation}

This rescaling of the uncertainty is used in the eclipse profile fitting (Section \ref{subsec:eclipse}), the spectroscopic phase curve analysis (Section \ref{subsec:spec_pc_analysis}), and the 4.2 $\upmu$m feature investigation (Section \ref{sec:CO2_asymm}).

\subsection{Characterization of the Total Eclipse}\label{subsec:eclipse}

\begin{deluxetable}{ll}
\tablecaption{Updated ZTF0038B binary parameters including new constraints from the eclipse profile fitting and eclipse timing analysis. For the eclipse profile fitting, a fixed prior is assumed from \cite{van_roestel_ztfj00382030_2021}. However, subsequent eclipse timing analysis provides an updated period that is consistent with the previously published value, but with much higher precision.  \label{tab:binary_params_tab}}
\tablewidth{0pt}
\tablehead{}

\startdata
\multicolumn{2}{l}{\textbf{Priors for Eclipse Fit}} \\
\hline
$t_0$ (MJD$_{\rm TDB}$)\tablenotemark{*} & $\mathcal{U}(60648.16,60648.18)$ \\
$P$ (days)\tablenotemark{**} & 0.43192080512$^\dagger$ \\
$R_{\rm WD}$ ($R_\odot$)\tablenotemark{*} & 0.01429$^\dagger$ \\
$R_{\rm BD}/R_{\rm WD}$\tablenotemark{*} & $\mathcal{U}(5.0,6.0)$ \\
Inclination ($^\circ$)\tablenotemark{*} & 89.71$^\dagger$ \\
Separation $a$ ($R_\odot$)\tablenotemark{*} & 1.987$^\dagger$ \\
$u$ (limb darkening coeff.) & $\mathcal{U}(-1, 1)$ \\
offset & $\mathcal{U}(0.9, 1.1)$ \\
\hline
\multicolumn{2}{l}{\textbf{Binary Parameters}} \\
\hline
$d$ (pc)\tablenotemark{*} & 139 ($\pm$2) \\
$P$ (days)\tablenotemark{***} & 0.43192080512 ($\pm$67) \\
$t_0$ (MJD$_{\rm TDB}$)\tablenotemark{**} & 60648.3433018 ($\pm$14) \\
$a$ ($R_\odot$)\tablenotemark{*} & 1.987 ( \iasym{30}{22} ) \\
$R_{\rm WD}$ ($R_\odot$)\tablenotemark{*} & 0.01429 ( \iasym{22}{17} ) \\
$R_{\rm BD}$ ($R_\odot$)\tablenotemark{**} & 0.0780 ($\pm$11) \\
$M_{\rm WD}$ ($M_\odot$)\tablenotemark{*} & 0.505 ( \iasym{24}{18} ) \\
$M_{\rm BD}$ ($M_\odot$)\tablenotemark{*} & 0.0593 (\iasym{36}{39}) \\
$u_1$\tablenotemark{**} & 0.292 ( \iasym{66}{70} ) \\
$C_\text{norm}$\tablenotemark{**} & 1.00403 ($\pm$68) \\
$\log g_{\rm WD}$ (cgs)\tablenotemark{*} & 7.832 ($\pm$13) \\
$\log g_{\rm BD}$ (cgs)\tablenotemark{*} & 5.425 ( \iasym{20}{30} ) \\
\enddata

\tablecomments{
Uncertainties in parentheses apply to the last digits shown (e.g. 7.832(13) = 7.832 ± 0.013).\\
$^\dagger$ Fixed prior values.
$\mathcal{U}$ denotes a uniform prior range.
}
\tablenotetext{*}{\quad Values from \cite{van_roestel_ztfj00382030_2021} \newline($t_{0}$ propagated to epoch of JWST observations).}
\tablenotetext{**}{\quad Parameters from \texttt{batman} eclipse fit.}
\tablenotetext{***}{\quad Period from eclipse timing analysis.}
\end{deluxetable}

Although the scatter in the high-cadence light curve is greater than the scatter of the light curve produced at the native time-sampling of the observation, the pipeline uncertainties do not account for this increased scatter. For this reason, the error inflation technique described in Section \ref{subsec:lc_production} is particularly important for our eclipse-fitting procedure, which utilizes the high-cadence time-series data introduced in Section \ref{subsec:high_cadence_extraction}.

\begin{figure*}[ht]
    \centering
    \includegraphics[width=1.0\linewidth]{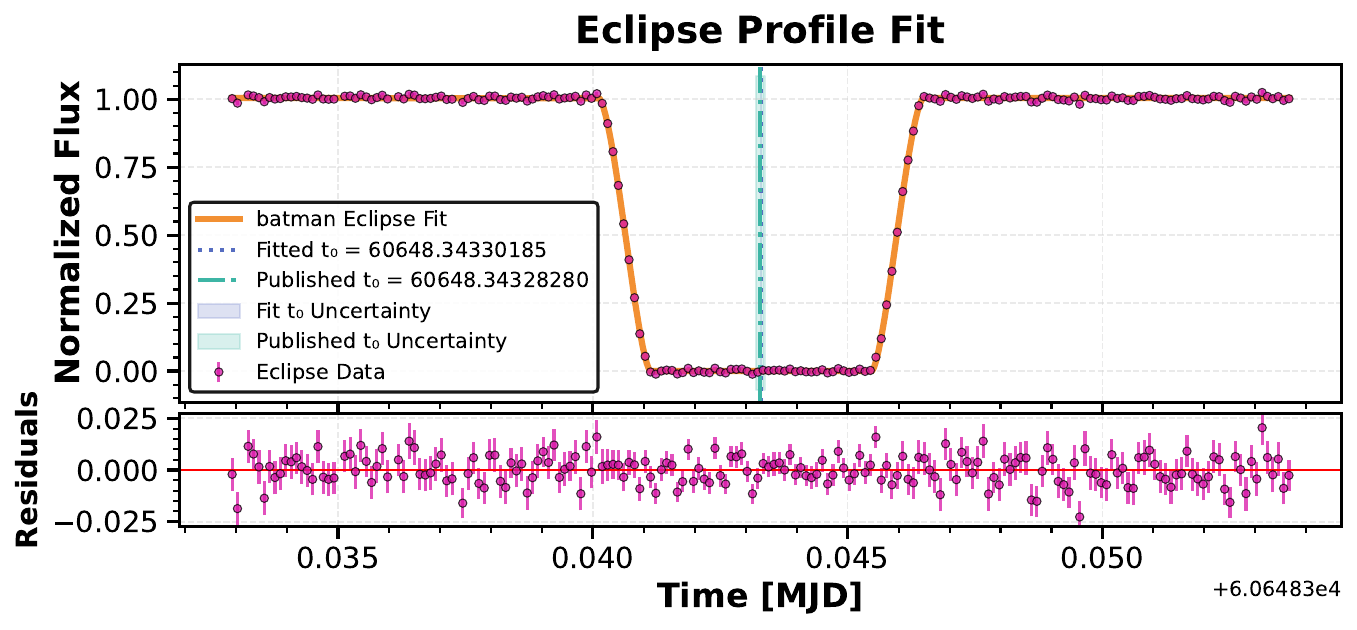}
    \caption{Best-fit \texttt{batman} transit model  of ZTF0038 fit via MCMC using \texttt{emcee}. The brown dwarf's emission is subtracted, and flux is normalized such that the out-of-eclipse signal is centered around one, and the flux goes to zero during the total eclipse.}
    \label{fig:eclipse_fit}
\end{figure*}

The eclipse profile is fit using the same time interval surrounding the eclipse as used to separate the brown dwarf and white dwarf signal (highlighted in pink in Figure~\ref{fig:isolating_BD}). As discussed in Section \ref{subsec:separating_WD_and_BD}, the brown dwarf's variation is treated as negligible during this 16.7 minute time interval. We subtract the brown dwarf emission contribution from the light curve and normalize the light curve to the median of the out-of-eclipse data. The resulting light curve (Figure \ref{fig:eclipse_fit}) is effectively an isolated white dwarf eclipse light curve without any contribution from the brown dwarf. Because the brown dwarf totally eclipses the white dwarf, the normalized flux reaches 0 during the transit. This satisfies the model assumptions present in the \texttt{batman} transit model package \citep{Kreidberg2015}, which is designed to fit exoplanet transits, and does not account for flux contributions from the companion. The light travel time between far-side (brown dwarf eclipse) and the near-side (brown dwarf transit) is 9.2 seconds. Our timing measurement uses the barycentric mid-transit time. This reference time needs no light travel time delay correction.

We determine the best-fit \texttt{batman} eclipse model using the Markov Chain Monte Carlo (MCMC) fitting algorithm implemented by the \texttt{emcee} package \cite{foreman-mackey_emcee_2013}. We use 2000 walkers with 10,000 steps per walker, discarding the first 5000 steps as burn-in. We adopt a fixed period, inclination, binary separation, and white dwarf radius from \cite{van_roestel_ztfj00382030_2021}. The free parameters in our model are the radius of the brown dwarf scaled by the white dwarf radius ($R_{\rm BD}/R_{\rm WD}$), the eclipse center ($t_0$), the limb darkening parameter(s) ($u_1$, $u_2$, etc.), and the normalization constant ($C_\text{norm}$). Four limb-darkening options are explored: uniform, linear, quadratic, and exponential. The log-likelihood is maximized for the linear and quadratic options with essentially indistinguishable residuals. The linear limb-darkening model is adopted because it is the simpler form. Linear limb darkening introduces one free parameter, $u_1$. The uniform priors and binary parameters are listed in Table~\ref{tab:binary_params_tab}. The best-fit eclipse model is shown in Figure~\ref{fig:eclipse_fit}, and the corner plot presenting the posterior distributions is shown in Appendix \ref{sec:eclipse_profile_appendix}.

\subsection{Eclipse Timing Analysis}

We use the period and ephemeris from \citet{van_roestel_ztfj00382030_2021} to predict the ephemeris at our observing epoch. Assuming a constant period, we propagate the ephemeris to the epoch of our JWST observations assuming that the mid-eclipse time occurs exactly $N_\text{orbits}$ after the published ephemeris:

\begin{equation}
    t_\text{0, propagated} = N_\text{orbits} \cdot P_\text{van Roestel}
\end{equation}

After 3711 orbits have occurred ($N_\text{orbits} = 3711 $) the uncertainty of the propagated ephemeris is:

\begin{equation}
    \sigma_{t_\text{0, propagated}} = \sqrt{\sigma_{t_\text{0, van Roestel}} + N_\text{orbits} \cdot \sigma_{P_\text{van Roestel}}},
\end{equation}

The propagated ephemeris is $t_\text{0, propagated} = 60648.34328 \pm 8.5\times10^{-5} 
\text{MJD}_\text{TDB}$.
The period and ephemeris from our eclipse timing analysis agree with this prediction to within $1\sigma$, indicating that the orbital period of ZTF0038 is highly stable over the baseline of 3711 orbits or $\sim 1600$ days.

We present an updated period for ZTF0038 using the mid-eclipse timing from \cite{van_roestel_ztfj00382030_2021}, our mid-eclipse time listed in Table \ref{tab:binary_params_tab}, and $N_\text{orbits} = 3711$. By assuming that the duration between the two eclipses covers exactly 3711 orbits, we calculate a period of $ 0.43192080512 \pm 6.7 \times 10^{-10}$ days. These two epochs of observations constrain the period to a precision of 60 $\upmu$s, which surpasses the native precision of the \textit{JWST} clock by two orders of magnitude \citep{shaw_calibrating_2025}.

\subsection{Broadband Phase Curve Analysis}\label{subsec:broadband_pc_analysis}

\begin{figure*}[!t]
    \centering
    \includegraphics[width=0.85\linewidth]{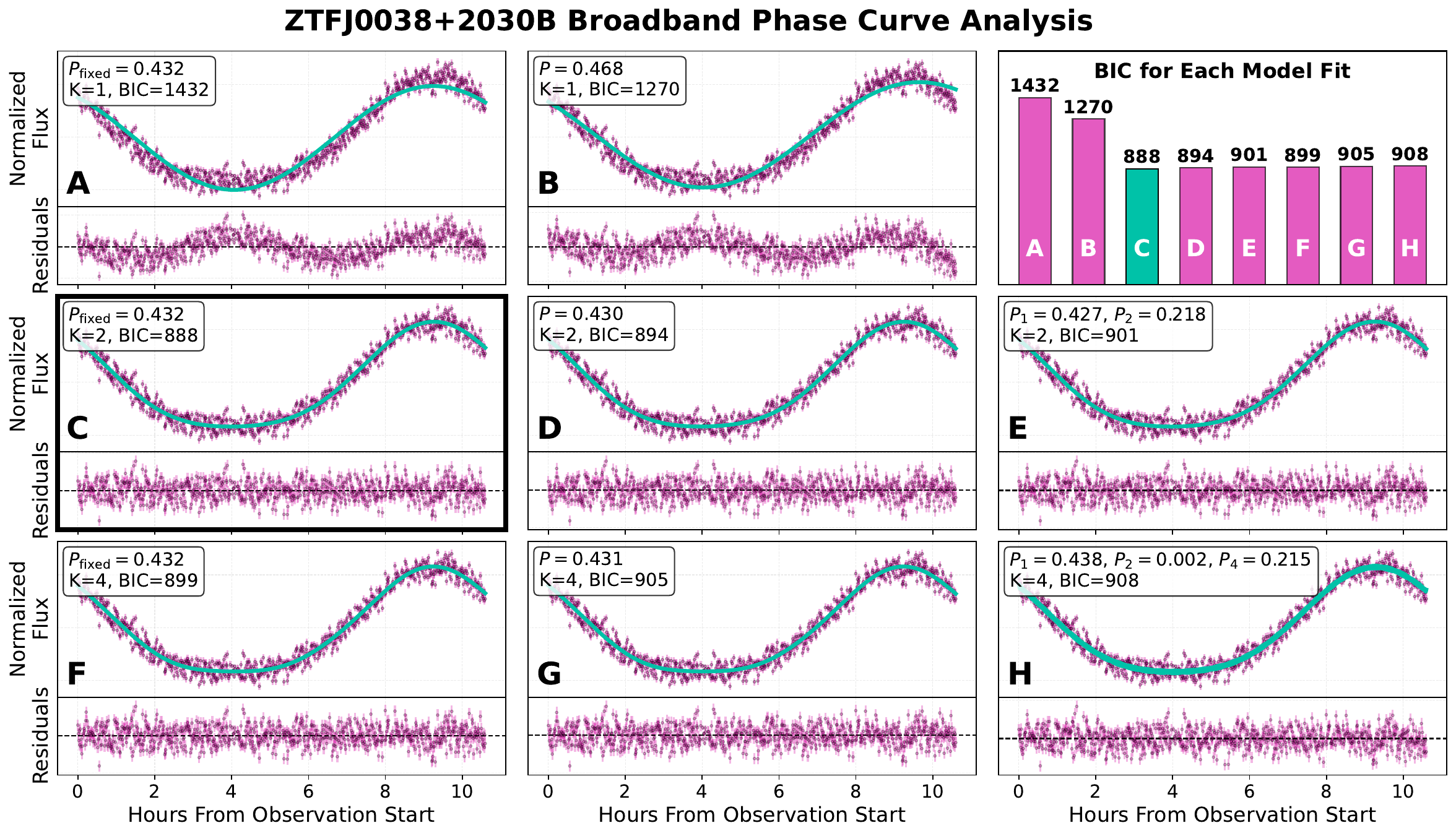}
    \caption{Several multi-component sinusoidal model fits of the broadband phase curve of ZTF0038B. The top row uses a single sinusoid to fit the data, and each row below adds an additional term. The first column fits the model assuming the period found in our eclipse timing analysis (Table \ref{tab:binary_params_tab}), the second column shares a single best-fit period among all terms, and the third column fits for the period of each term separately. The bar chart on the top right compares the BIC value from each model fit, and concludes that panel C provides the optimal phase curve model. This consists of a two-component fit that is consistent with the period determined from our eclipse timing analysis.} \label{fig:broadband_phase_curve}
\end{figure*}

We explore eight models of increasing complexity to fit the broadband phase curve of ZTF0038B. Because simpler models are nested within more complex ones, we use the Bayesian Information Criterion (BIC) to select the optimal model complexity. The general model form (Equation \ref{eqn:pc_model}) resembles a modified Fourier series expansion. The third harmonic is excluded due to the symmetry in hemispherically integrated observations \citep{cowan_odd_2017}, and the period is not necessarily shared by all components in the series. The pixel-level uncertainties are combined in quadrature

\begin{equation} \label{eqn:pc_model}
    A_0 + \sum_{\substack{k=1 \\ k \ne 3}}^{k_\text{max}} 
    A_k \cdot \sin\!\left(\frac{2\pi k t}{P_k} - \phi_k\right)
\end{equation}

We fit this model up to $k=4$, excluding $k=3$ \citep{cowan_odd_2017}. We explore three assumptions about the period: (1) A fixed period found from our eclipse timing analysis summarized in Table \ref{tab:binary_params_tab} ($P_k = P_{\text{fixed}}=0.43192080512$ days). (2) A single best-fit period ($P_k=P$) fit as one free parameter that is present in all components. (3) Individual best-fit periods fit independently for each sinusoidal component ($P_k$). The model fits and their associated BIC values are produced using \texttt{lmfit} \citep{newville_2025_16175987}, and presented in Figure \ref{fig:broadband_phase_curve} as well as the supplementary data behind Figure \ref{fig:amplitude_phase_vs_wavelength}. Lower BIC values indicate better fits that balance goodness-of-fit against model complexity. We find that the $k_\text{max}=2$ model with fixed period is favored. The best-fitting phase curve period is therefore consistent with the orbital period determined by our eclipse timing analysis.

\subsection{Spectroscopic Phase Curve Analysis}\label{subsec:spec_pc_analysis}

\begin{figure*}
    \centering
    \includegraphics[width=1.0\linewidth]{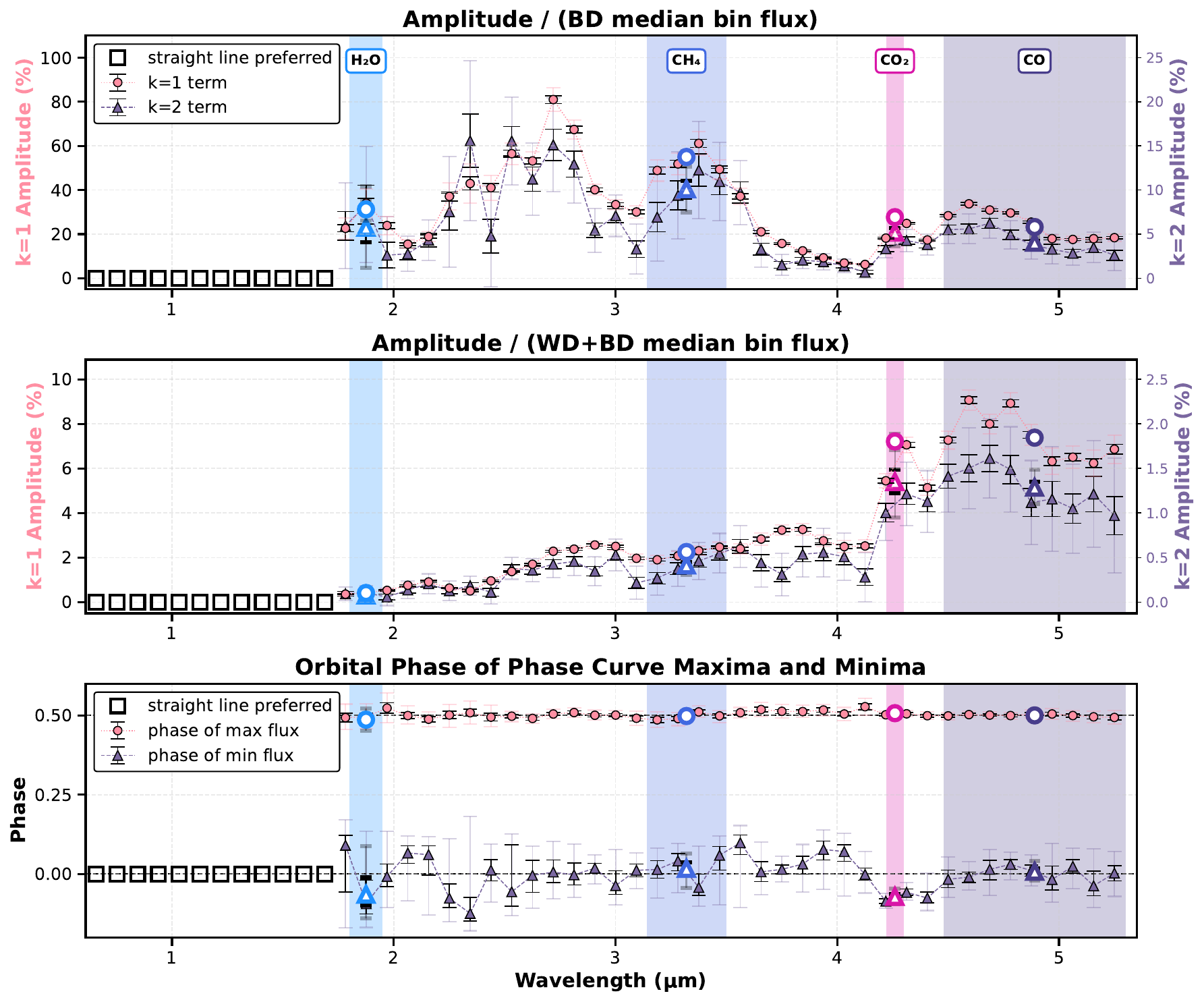}
    \caption{
    Fourier amplitudes and phases of the spectroscopic phase curves as a function of wavelength. Each phase curve is fit with a second-order Fourier series; bins consistent with a flat model are indicated with black squares. In each panel, we include the phase curve information for 50 evenly spaced wavelength bins as well as a few additional wavelength bins corresponding to H$_2$O, CH$_4$, CO$_2$, and CO absorption features. The wavelength ranges used to make these phase curves are highlighted. The markers for these molecular absorption phase curves are bolded. All data points include black error bars corresponding to the 1-$\sigma$ uncertainties, and fainter gray error bars that correspond to the 3-$\sigma$ uncertainties. \textbf{Top:} Phase curve amplitudes normalized by the brown dwarf flux. The $k=1$ (first-order) and $k=2$ (second-order) components are shown with circular and triangular markers, respectively. \textbf{Center:} The same as the top panel, except the amplitudes are normalized by the combined flux of the white dwarf and brown dwarf. \textbf{Bottom:} The phases corresponding to the maximum and minimum points in the phase curve model are shown with circular and triangular markers, respectively.}
    \label{fig:amplitude_phase_vs_wavelength}
\end{figure*}

We determine whether modulation is detected in each of the 50 spectroscopic phase curves by comparing each bin with a flat line. For all wavelength bins with modulation detected, we fit the phase curve with our optimal model found in our broadband phase curve analysis: a second-order Fourier series with the periods of both components fixed at the system period.  The resulting amplitudes of the $k=1$ and $k=2$ Fourier terms as well as the phase of the maximum and minimum points in each model phase curve are shown in Figure~\ref{fig:amplitude_phase_vs_wavelength}, and a selection of spectroscopic phase curves with their best-fit models is shown in Figure~\ref{fig:spec_pcs}. 

The uncertainty on the first and second order amplitudes shown in Figure \ref{fig:amplitude_phase_vs_wavelength} are the standard $1\sigma$ errors produced by \texttt{lmfit} assuming Gaussian noise. In Figure \ref{fig:amplitude_phase_vs_wavelength}, we also present the phase that corresponds to the minimum and maximum flux in the phase curve models. This information is not parameterized in our phase curve model, so we use bootstrap resampling with replacement to estimate uncertainties. For 1200 draws per bin, we resample the phase curve data and fit the resampled data with the Fourier series model. The phase of maximum and minimum flux is recorded for each resampling, and the error bars are generated from the resulting distribution of these parameters for the 1200 draws. The nominal $1\sigma$ uncertainties are determined using the central 16th–84th percentiles, and the  central 0.135th–99.865th percentiles are used for the 3$\sigma$ uncertainties.

For each wavelength bin, we compute the $\chi^2$ statistic for a flat model fit to evaluate whether a significant modulation signal is present. Each phase curve comprises 820 flux measurements, and the flat model has one free parameter, giving 819 degrees of freedom. We reject the flat model and consider variability detected when the $\chi^2$ statistic exceeds the threshold corresponding to a $3\sigma$ confidence level for this distribution. Phase curves consistent with a flat model are indicated with black squares in Figure~\ref{fig:amplitude_phase_vs_wavelength}.

Beyond 1.7~$\upmu$m, significant ($>3\sigma$) modulation is consistently detected in each wavelength bin. The first- and second-order Fourier amplitudes are tightly correlated across wavelength (Figure~\ref{fig:amplitude_phase_vs_wavelength}, top panel). The overall modulation amplitude increases with wavelength as the white dwarf flux decreases steeply at the long wavelength end (Figure~\ref{fig:amplitude_phase_vs_wavelength}, middle panel). The phase curve maximum in each bin is consistent with the substellar point (phase = 0.5) within 3-$\sigma$. The phase curve minimum is consistent with the anti-stellar point (phase = 0) for all bins except for a notable deviation around 4.2~$\upmu$m (Figure~\ref{fig:amplitude_phase_vs_wavelength}, bottom panel).

\section{Spectroscopic Analysis:}\label{sec:spectroscopic_analysis}

\subsection{Representative Phase-Resolved Spectra}

\begin{figure*}
    \centering
    \includegraphics[width=0.8\linewidth]{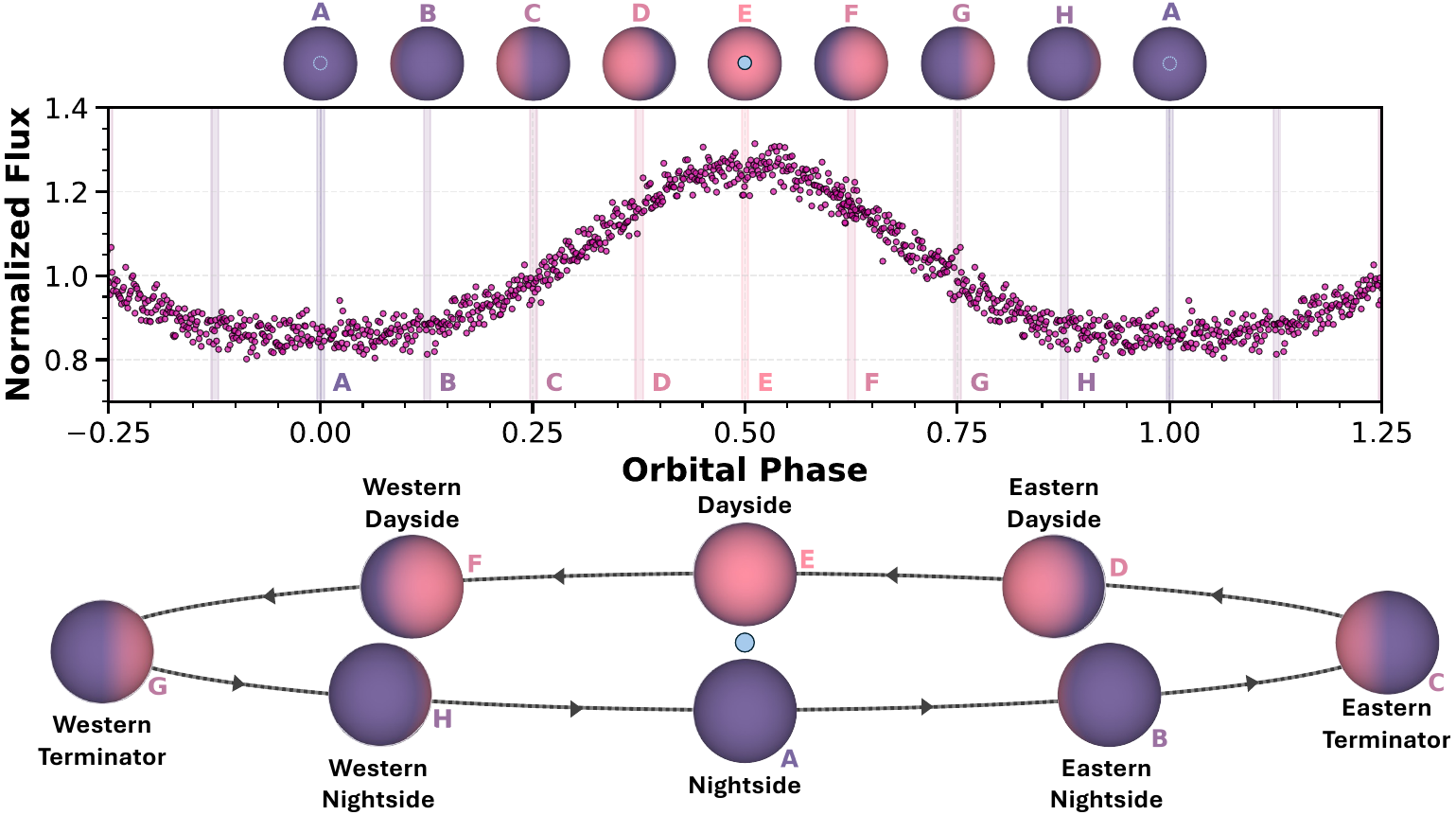}
    \caption{Illustration of the WD--BD geometry at eight representative phases in the orbit. The upper illustrations visualize a simple representation of the brown dwarf hemisphere observed at each phase. The shading indicates
how much of the dayside (pink) and nightside hemisphere (purple) are visible at each phase. In phases A and E (nightside and dayside), the smaller central circle indicates when the white dwarf is in behind or in front of the brown dwarf. The center panel shows the phase-folded, broadband phase curve. The highlights indicate the amount of time used to produce each median-combined spectrum for each of the eight representative phases. Each of the eight highlighted phases is also shown in the lower orbital diagram, where the inclination is adjusted to show all phases of the orbit more clearly.}
    \label{fig:phase_illustration}
\end{figure*}

\begin{figure*}
    \centering
    \includegraphics[width=1.0\linewidth]{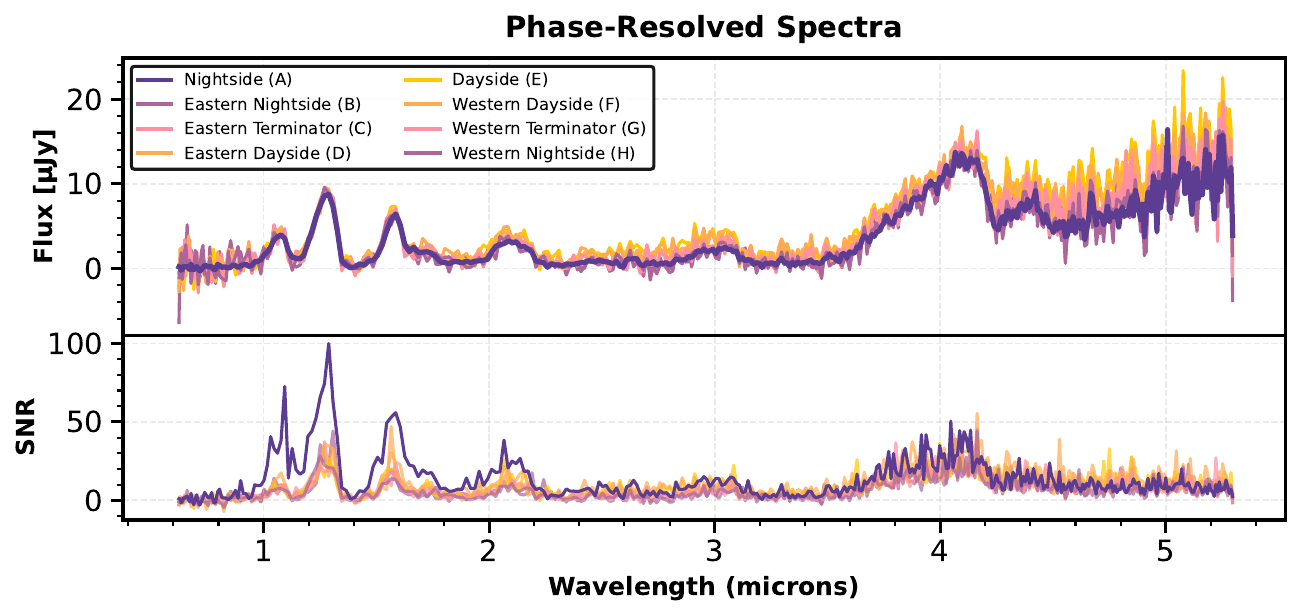}
    \caption{Emission spectra of ZTF0038B at 8 representative phases (top panel) and their associated signal-to-noise ratios (bottom panel).}
    \label{fig:SNR_fig}
\end{figure*}

We generate representative spectra for the dayside and nightside as well as 6 intermediate phases for model comparison and atmospheric retrievals (shown in Figure \ref{fig:phase_illustration}). From the 832 spectra observed throughout an entire rotation of ZTF0038, we generate high SNR spectra to represent 8 phases in the brown dwarf's rotation: the dayside, nightside, the eastern and western terminator, and four intermediate phases. The phases are determined by dividing the period listed in Table \ref{tab:binary_params_tab} into 8 evenly spaced intervals. The nightside phase corresponds to phase=0, and is centered around the mid-eclipse time ($t_0$). All other phases are determined relative to this nightside phase. We use the duration of the total eclipse as the approximate window of time to generate a high signal-to-noise spectrum centered around each of the 8 phases. This duration of $\sim6$ minutes centered around each of the 8 `representative' phases corresponds to 0.85\% of the period. As discussed in Section \ref{subsec:separating_WD_and_BD}, the brown dwarf's variability during the duration of the eclipse introduces negligible uncertainties compared to the per-integration flux uncertainty. The uncertainty of the median-combined spectra are calculated using Equation \ref{eqn:median_combined_err}. This empirical estimation of the uncertainty ensures that the analysis is not impacted by the wavelength-dependent systematics presented in Figure \ref{fig:spec_pcs}.

Figure \ref{fig:SNR_fig} shows the spectroscopic signal-to-noise ratios (SNR) of the 8 representative spectra. These are calculated using our empirical estimation of the uncertainty (Equation \ref{eqn:median_combined_err}). The maximum SNR is achieved in the \textit{J} band and around the 4 $\upmu$m continuum. The spectrum with the highest SNR is the nightside spectrum, which was observed fully during the total eclipse, and it reaches an SNR of $\sim80$ in the \textit{J} band, and $\sim50$ in the longer wavelengths. All other phases achieve a maximum SNR of $\sim40-50$.

\subsection{Comparison to Isolated Brown Dwarf Observations}\label{subsubsec:comp_isolated_BD}

\begin{figure*}[ht]
    \centering
    \includegraphics[width=1\linewidth]{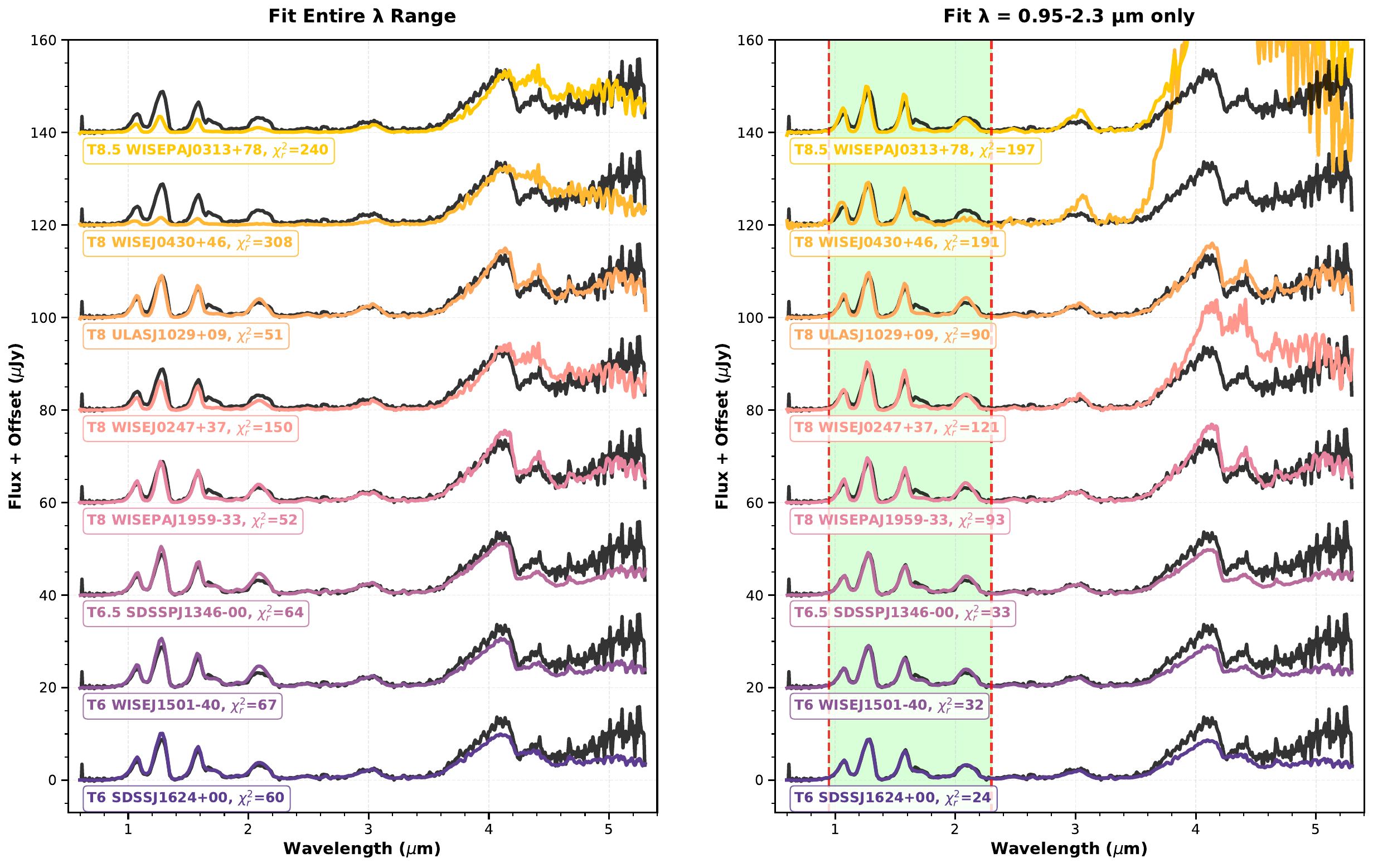}
    \caption{Comparison of the observed ZTF0038B spectrum (black) with archival \textit{JWST} PRISM spectra of mid-to-late T dwarfs \citep{beiler_precise_2024}. We uniformly scale the archival spectra to minimize the chi-squared statistic across the entire wavelength range (left) and across the \textit{J}, \textit{H}, and \textit{K} bands only (right).}
    \label{fig:field_bd_comparison}
\end{figure*}

The nightside emission resembles that of isolated, non-irradiated brown dwarfs. In Figure \ref{fig:field_bd_comparison}, we compare our nightside spectrum with several mid-to-late T dwarfs \citep{beiler_precise_2024}. We use isolated brown dwarf spectra as templates and determine the scaling factor that minimizes $\chi^2_r$ for each template fit to ZTF0038B's nightside spectrum. The left panel of Figure \ref{fig:field_bd_comparison} shows that ULASJ1029+09 (T8) provides the best overall fit. However, brown dwarf spectral types are typically classified using only the \textit{J}, \textit{H}, and \textit{K} bands. We therefore identify the best template match within these bands. We find that SDSSJ1624+00 (T6) provides the closest fit in the \textit{J}, \textit{H}, and \textit{K}  bands, though it shows less emission at longer wavelengths than ZTF0038B. In 1999, SDSSJ1624+00 was the first free-floating T-dwarf discovered \citep{strauss_discovery_1999}, and has since been adopted as a standard reference spectrum for T6 dwarfs \citep{burgasser_unified_2006}.

\subsection{Brightness Temperature}\label{subsec:brightness_temp}

\begin{figure*}[!th]
    \centering
    \includegraphics[width=0.75\linewidth]{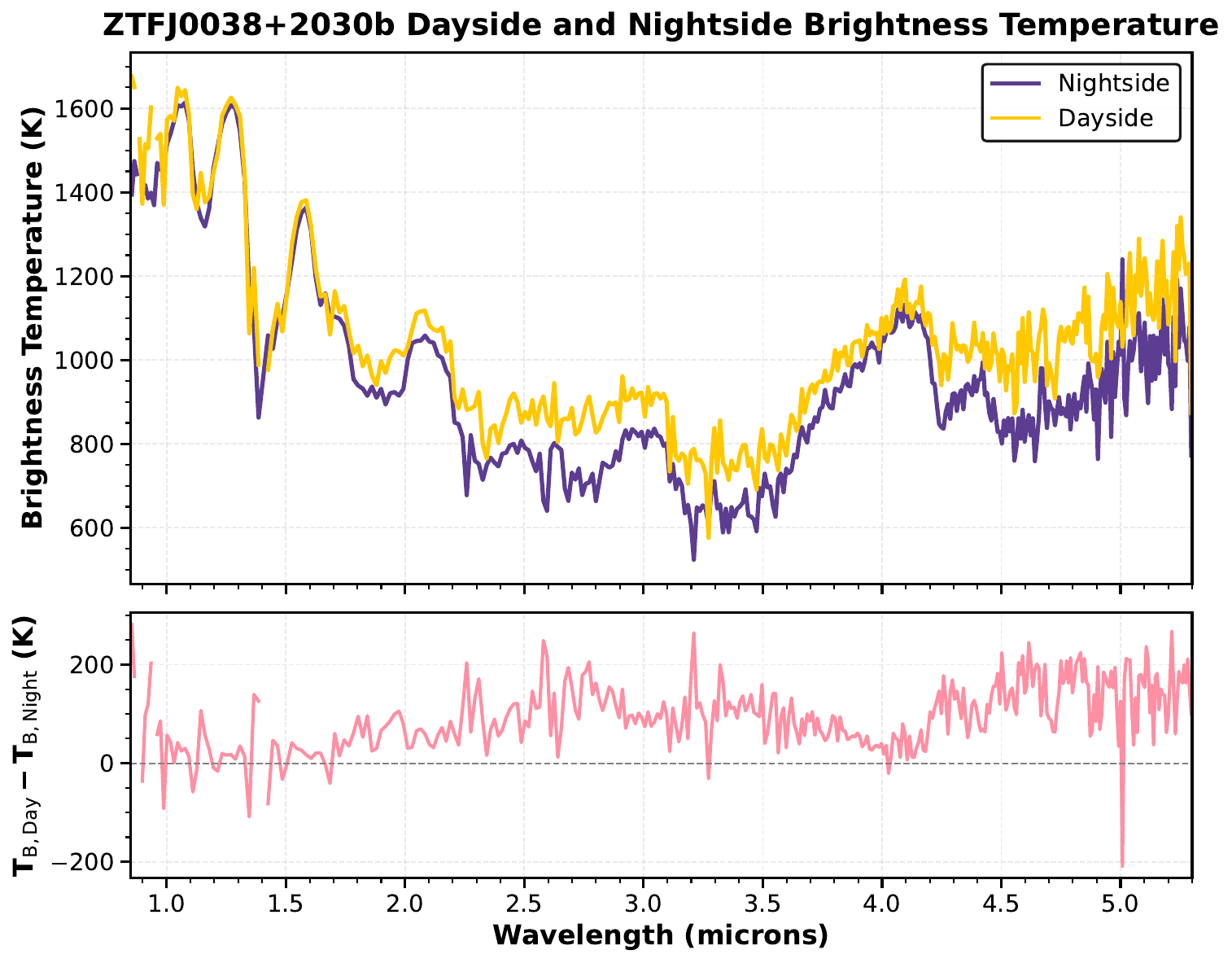}
    \caption{\textbf{Top:} ZTF0038B dayside and nightside brightness temperature. \textbf{Bottom:} Difference between the dayside and nightside brightness temperatures.}
    \label{fig:brightness_temp}
\end{figure*}

In Figure \ref{fig:brightness_temp}, we present the spectroscopic brightness temperature of the dayside and nightside emission of ZTF0038B. We calculate these brightness temperatures by first converting the observed flux $F_\nu$ to specific intensity $B_\nu$ using the brown dwarf radius and distance from Table \ref{tab:binary_params_tab}:
\begin{equation}
    B_\nu = \frac{F_\nu}{\pi \cdot (R_\text{BD}/d_\text{BD})^2} 
\end{equation}
where all quantities are in cgs units. We then invert the Planck function 
\begin{equation}
 B_\nu(T_B) = \frac{2h\nu^3}{c^2} \frac{1}{\exp(h\nu/k_BT_B) - 1}
\end{equation}
to obtain the brightness temperature $T_B$:
\begin{equation}
 T_B = \frac{h\nu}{k_B \ln\left(\frac{2h\nu^3}{c^2 B_\nu} + 1\right)}.
\end{equation}

In the lower panel of Figure \ref{fig:brightness_temp}, we also present the difference between the dayside and nightside brightness temperatures of ZTF0038B.

\subsection{Forward Modeling}\label{subsec:forward_modeling}

\begin{deluxetable*}{lcc|lc}
\tablecaption{Best-fit parameters for the nightside and dayside emission spectra of ZTF0038B.
\label{tab:grid_fits}}

\tablewidth{0pt}
\tablehead{
\multicolumn{3}{c|}{\textbf{Sonora Elf Owl (Nightside)}} & \multicolumn{2}{c}{\textbf{Irradiated Grid (Dayside)}}
}
\startdata
\hline
$T_\text{eff}$ [K]              & $\mathcal{U}[900, 1000]$        & $974.3^{+3.3}_{-3.2}$          & $T_\text{WD}$ [K]$^\dagger$            & 11000   \\
log($g$) [cgs]                  & $\mathcal{N}(5.425, 0.025)$     & $5.383^{+0.016}_{-0.016}$      & log($g_\text{WD}$) [cgs]$^\dagger$     & 7.75    \\
$M_{BD}$ [M$_\text{Jup}$]       & $\mathcal{N}(62.09, 3.93)$      & $57.6^{+2.0}_{-2.2}$           & $M_{BD}$ [M$_\text{Jup}$]$^\dagger$    & 62.121  \\
$R_{BD}$ [R$_\text{Jup}$]       & $\mathcal{N}(0.762, 0.012)$     & $0.7683^{+0.0059}_{-0.0064}$   & $R_{BD}$ [R$_\text{Jup}$]$^\dagger$    & 0.762   \\
$\omega$ [mas]                  & $\mathcal{N}(7.19, 0.11)$       & $7.223^{+0.060}_{-0.056}$      & $R_{WD}$ [R$_\odot$]$^\dagger$         & 0.01429 \\
log($L_{BD}/\text{L}_\odot$)           & --                              & $-5.2956^{+0.0077}_{-0.0076}$  & $a$ [AU]$^\dagger$                     & 0.009   \\
Fe/H                            & $\mathcal{U}[-1.0, 1.0]$                              & $-0.150^{+0.011}_{-0.011}$     & Age [Gyr]                              & 10      \\
C/O                             & $\mathcal{U}[0.23, 1.145]$        & $0.3958^{+0.0071}_{-0.0073}$   & M/H                                    & 0       \\
log($k_{zz}$)                   & $\mathcal{U}[2.0, 9]$                              & $2.015^{+0.025}_{-0.011}$      & $f_\text{sed}$                         & 1       \\
\enddata
\tablecomments{The nightside is fit with the Sonora Elf Owl grid \citep{Mukherjee2024} using nested sampling. The priors are shown in the second column, and the parameters of the best-fit model are shown in the third column. Uniform and Gaussian priors are denoted $\mathcal{U}$ and $\mathcal{N}$, respectively. The dayside is fit with an irradiated substellar atmosphere forward model grid discussed in Section \ref{subsec:forward_modeling}. The parameters of the best-fitting model grid point are listed in the final column, and fixed prior values are indicated with $\dagger$.}
\end{deluxetable*}

We utilize forward models representing both the non-irradiated nightside and irradiated dayside hemispheres of the brown dwarf. For the nightside, we adopt the Sonora Elf Owl disequilibrium chemistry grid \citep{Mukherjee2024}. This is an existing, cloud-free, self-consistent atmospheric model grid with publicly available pre-calculated synthetic spectra. The Elf Owl grid includes models with C/O between 0.23 -- 1.145 and Fe/H between -1.0 -- 1.0. Cloud-free models are appropriate for this T-dwarf because the atmosphere is typically cool enough that clouds settle below the photosphere and do not significantly impact the spectrum. The nightside spectrum of ZTF0038 resembles that of field brown dwarfs, justifying the use of self-luminous atmosphere models. 

The Sonora Elf Owl grid is incorporated into \texttt{species}, a toolkit that interpolates within model parameter space and uses Bayesian nested sampling to produce posterior distributions for the fitted parameters \citep{stolker_miracles_2020}. We use the \texttt{ultranest} Bayesian inference package \citep{buchner_ultranest_2021} to produce posterior distributions. We adopt normal priors for the mass ($M$), surface gravity ($\log g$), radius ($R_{BD}$), and parallax ($\omega$) using the published values and uncertainties from \cite{van_roestel_ztfj00382030_2021}. The priors, posteriors, and best-fit parameters are listed in Table \ref{tab:grid_fits}. The corner plot and best-fit model spectrum are shown in Figure \ref{fig:elf_owl_corner_plot}. 

The Sonora Elf Owl model reproduces the emission features between $\sim0.6 - 4 \upmu$m, but struggles substantially at the longer wavelengths covered with NIRSpec PRISM. This wavelength regime contains a strong CO$_2$ feature, a strong CO feature, and a significant excess beyond $\sim4.5$ $\upmu$m. The fundamental brown dwarf parameters (mass, surface gravity, radius, parallax) are generally in agreement with published values from \cite{van_roestel_ztfj00382030_2021}.

We carried out an additional Elf Owl model fit for only the longer wavelengths (beyond 3.3 $\upmu$m), and while we found that the model could reproduce the CO$_2$ absorption at 4.2 $\upmu$m,  the best-fitting parameters did not converge in the corner plot, and they were highly discrepant with previously published values for the brown dwarf. This indicates that there is additional physics shaping the atmosphere of ZTF0038B that is not included in the Sonora Elf Owl model, which is constructed to fit non-irradiated, isolated brown dwarfs.

We verified that a cloud-free model was preferred by comparing the Elf Owl fit with the Sonora Diamondback grid \citep{morley_sonora_2024}, which includes clouds and assumes chemical and radiative-convective equilibrium. The best-fit Diamondback model preferred a large f$_\text{sed}$. However, none of the Diamondback models fit the nightside spectrum as well as the cloudless Elf Owl model. This result justifies our preference of a cloud-free model that incorporates disequilibrium chemistry.

To fit the dayside spectrum, we construct a grid of irradiated atmosphere models. With EGP, the 1D radiative-convective equilibrium model of \citet{Marley1999, Ackerman2001}, we ran a small grid of forward models for ZTF0038B. We varied the age of the system from 5, 8, and 10 Gyr, which determined the internal heat flux of the brown dwarf \citep{Marley2021}, the metallicity from Solar to 5x Solar, and explored various cloud sedimentation efficiencies, $f_\mathrm{fsed}$ = [5, 3, 2, 1] and cloud-free models. In the cloudy models, Na$_2$S, KCl, and ZnS clouds were allowed to condense. These forward models assume chemical equilibrium and that TiO and VO are sequestered at depth and not radiatively active in the observable atmospheres. We follow the methodology of prior works \citep{Amaro2023, Lew2022, French2024} to include the flux from the white dwarf host that falls outside of the wavelength range modeled by EGP in the same way, whereby we increased the irradiation in the first few wavelength bins to mimic holding the ultraviolet optical properties constant.

\begin{figure*}
    \centering
    \includegraphics[width=1.0\linewidth]{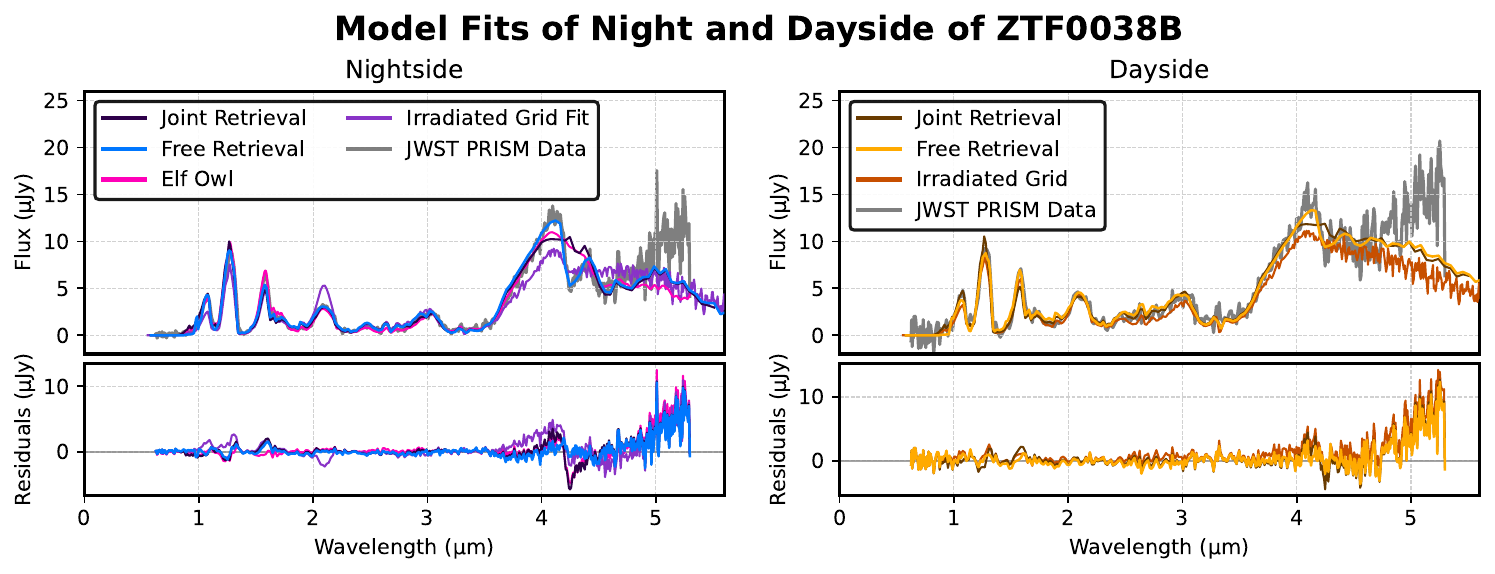}
    \caption{Nightside (left) and dayside (right) spectra of ZTF0038B compared with the best-fitting models obtained using forward-modeling grids and atmospheric retrievals. }
    \label{fig:model_fits}
\end{figure*}

To approximate the different hemispheres of the brown dwarf, we varied the recirculation efficiency parameter, ``rfacv", to approximate a nightside (0) or dayside (1) hemisphere. Finally, we compute the observed spectrum for the applicable observing geometry using \texttt{PICASO} \citep{Batalha2019paper} and match the resolution and wavelength range of the NIRSpec PRISM observation. The forward model fits and residuals for the dayside and nightside spectra are shown in Figure \ref{fig:model_fits}, and the best-fitting parameters are listed in Table \ref{tab:grid_fits}.  Similar to the Sonora Elf Owl model fit for the nightside, the irradiated grid model struggles to fit the absorption features at longer wavelengths. We emphasize that the irradiated model fitting for the dayside does not interpolate within the parameter space, so the best-fit parameters are limited to discrete grid points, and therefore do not include uncertainties.

\subsection{Joint, Chemically-Constrained Retrieval}\label{subsec:joint_retrieval}

\begin{figure*}[h]
    \centering
    \includegraphics[width=0.9\linewidth]{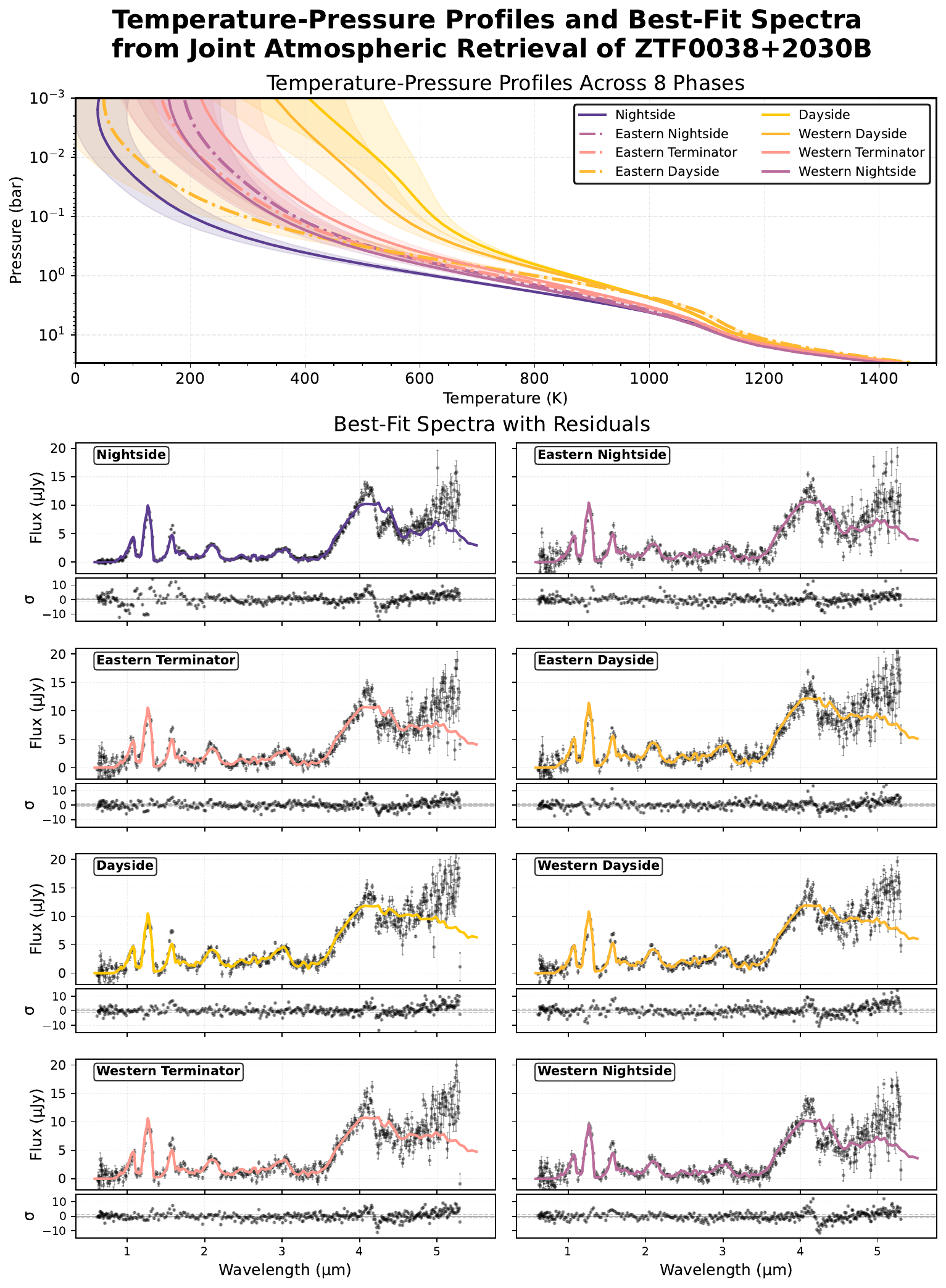}
    \caption{Joint multi-phase spectroscopic retrieval results. Upper panel: the retrieved temperature-pressure profiles for all eight phases. Lower panels: the observed and retrieved spectra for eight representative phases, as well as the residuals.}
    \label{fig:retrieval_fig}
\end{figure*}

We performed atmospheric retrievals using the \texttt{petitRadTrans} radiative transfer framework \citep{Molliere2019}. The retrieval employs nested sampling for parameter space exploration \citep{Feroz2009}. We adopted the temperature-pressure profile parameterization from \cite{Molliere2020}, which consists of three components: a deep atmosphere characterized by an internal temperature ($T_{\rm int}$), a middle atmosphere following the Eddington approximation, and an upper atmosphere described by a three-point spline. Chemical abundances were computed assuming equilibrium chemistry governed by bulk metallicity ([M/H]) and carbon-to-oxygen ratio (C/O), with additional flexibility through freely fit quench pressures that allow molecular abundances to deviate from equilibrium values at lower pressures \citep[e.g.,][]{Visscher:2011}.

Previous phase curve retrievals have generally relied on fits to spectra at individual phases \citep{Stevenson2014,Ashtari:2026}. If data quality is high enough, however, previous work has shown the efficacy of fitting spectroscopic phase curves with multiple longitudes fit simultaneously \citep{feng_2d_2020,Mikal-Evans:2022,Lothringer2024}, though usually with a linear combination of a dayside and nightside atmosphere. The high-quality phase-resolved spectra of ZTF0038B enabled a novel joint retrieval approach that fits longitude-specific TP profiles to each of the eight representative orbital phases simultaneously while sharing global parameters, including surface gravity ($\log g$), planetary radius, internal temperature ($T_{\rm int}$), C/O ratio, and bulk metallicity ([M/H]). However, the atmospheric temperature-pressure profiles and quench pressures were allowed to vary independently at each longitude. This physically motivated configuration captures longitudinal variations in atmospheric structure while constraining the fundamental planetary properties with the complete data set.

\begin{figure*}
    \centering
    \includegraphics[width=1\linewidth]{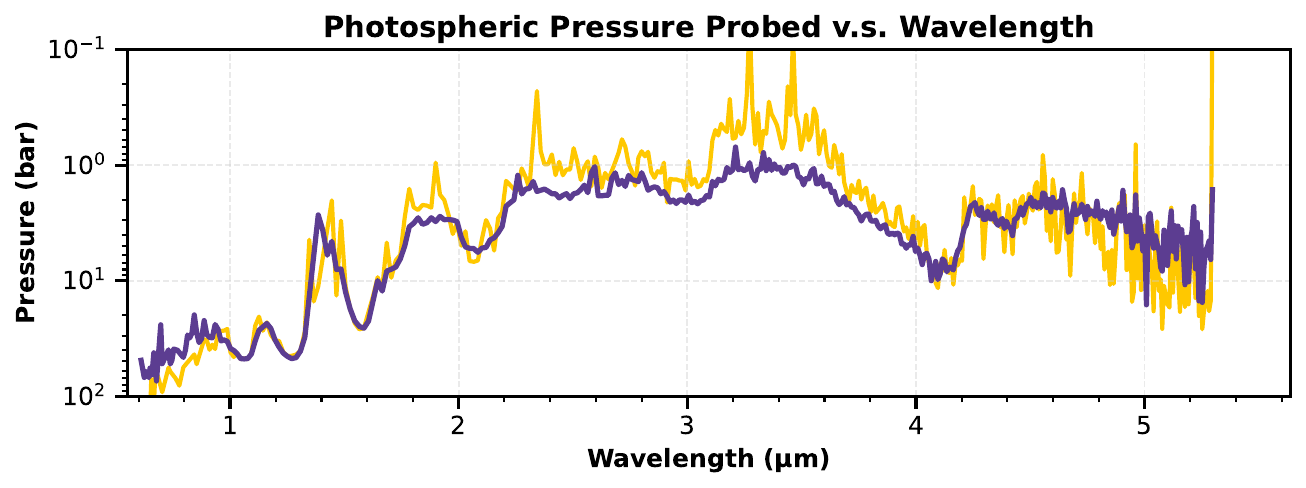}
    \caption{Dayside and nightside photospheric pressure probed as a function of wavelength for the ZTF0038B. We use the brightness temperatures presented in Section \ref{subsec:brightness_temp} and the temperature-pressure profiles from our joint retrieval to map the brightness temperature to its associated photospheric pressure value at each wavelength. }
    \label{fig:pressure_vs_wavelength}
\end{figure*}

The retrieved dayside and nightside spectra are shown in comparison with the forward modeling grids in Figure \ref{fig:model_fits}. The best-fit spectra for all 8 phases and their associated temperature-pressure profiles are shown in Figure \ref{fig:retrieval_fig}. The best-fit parameters are presented in Table~\ref{tab:retrieval_params}. In Figure \ref{fig:pressure_vs_wavelength}, we map the brightness temperatures calculated in Section \ref{subsec:brightness_temp} to their associated photospheric pressures using the temperature-pressure profile from the joint retrieval.

The joint retrieval successfully reproduced the observed spectra across all phases, except in the $\lambda > 4$\,$\upmu$m region, where CO$_2$ absorption shapes the spectrum. The retrieval also struggles to fit CO$_2$ even if the phases are fit independently. Global parameters were well-constrained by the joint analysis. The molecular abundances of H$_2$O, CO, CO$_2$, and CH$_4$ show modest variations with longitude, primarily driven by the differing temperature structures at each phase rather than changes in the underlying bulk chemistry. The shared metallicity and C/O ratio provided adequate fits across all phases, suggesting that the observed spectral variations are predominantly controlled by temperature-driven changes in molecular chemistry rather than fundamental compositional differences between hemispheres.

\begin{deluxetable*}{lccccccc}
\tablewidth{0pt}
\tablecaption{Best-fit parameters for the joint retrieval. \label{tab:retrieval_params}}
\tablehead{
    \multicolumn{8}{l}{\textbf{Joint Fit Parameters}} \\
    \colhead{$L_{\rm int}$ [$\log_{10}{(\frac{L}L_\odot)}$} &
    \colhead{$T_{\rm int}$ [K]} &
    \colhead{$\log(g)$ [cgs]} &
    \colhead{$R_{\rm BD}$ [R$_\text{Jup}$]} &
    \colhead{$M_{\rm BD}$ [M$_\text{Jup}$]} &
    \colhead{$M/H$} &
    \colhead{C/O} &
    \colhead{}
}
\startdata
-4.88 & $1246.0 \pm 2.8$ & $5.4415 \pm 0.0097$ & $0.7985 \pm 0.0051$ & $71.1 \pm 1.9$ & $-0.056 \pm 0.014$ & $0.563 \pm 0.011$ & \\
\hline
\hline
\multicolumn{8}{l}{\textbf{Phase-Dependent Parameters}} \\
\hline
\colhead{Phase} & \colhead{$T_{\rm 1bar}$ [K]} & \colhead{$\log_{10}(P_{\rm quench}) [bar]$ } & \colhead{$L$ [$\log_{10}{(\frac{L}L_\odot)}$]} & \colhead{$T_{\rm eff}$ [K]} & \colhead{$\chi^2$} & \colhead{Error Factor} & \colhead{} \\
\hline
Nightside (A)     & $647.3^{+17.2}_{-16.7}$& $1.46 \pm 0.03$ & $-5.26$ & $921.6$ & $15.912$ & $3.969 \pm 0.089$ & \\
E. Nightside (B)  & $740.4^{+13.3}_{-13.6}$ & $1.47 \pm 0.03$& $-5.22$ & $940.6$ & $8.604$  & $2.934 \pm 0.066$ & \\
E. Terminator (C) & $755.9^{+14.1}_{-14.1}$ & $1.35 \pm 0.03$ & $-5.21$ & $946.8$ & $7.253$  & $2.701 \pm 0.060$ & \\
E. Dayside (D)    & $824.0^{+13.8}_{-13.6}$ & $1.252 \pm 0.05$ & $-5.14$ & $987.9$ & $8.452$  & $2.910 \pm 0.063$ & \\
Dayside (E)       & $881.3^{+5.9}_{-5.7}$ & $0.87 \pm 0.24$ & $-5.15$ & $981.6$ & $8.924$  & $2.995 \pm 0.067$ & \\
W. Dayside (F)    & $867.1^{+6.6}_{-6.5}$ & $1.248 \pm 0.04$ & $-5.15$ & $983.7$ & $12.281$ & $3.519 \pm 0.078$ & \\
W. Terminator (G) & $786.3^{+11.6}_{-11.8}$ & $1.34 \pm 0.03$ & $-5.20$ & $953.1$ & $8.039$  & $2.848 \pm 0.063$ & \\
W. Nightside (H)  & $713.3^{+13.8}_{-13.7}$ & $1.46 \pm 0.03$ & $-5.26$ & $920.9$ & $8.430$  & $2.927 \pm 0.062$ & \\
\enddata 
\end{deluxetable*}

\subsection{Free Retrievals}\label{subsec:free_retrievals}

\begin{figure*}
    \centering
    \includegraphics[width=1\linewidth]{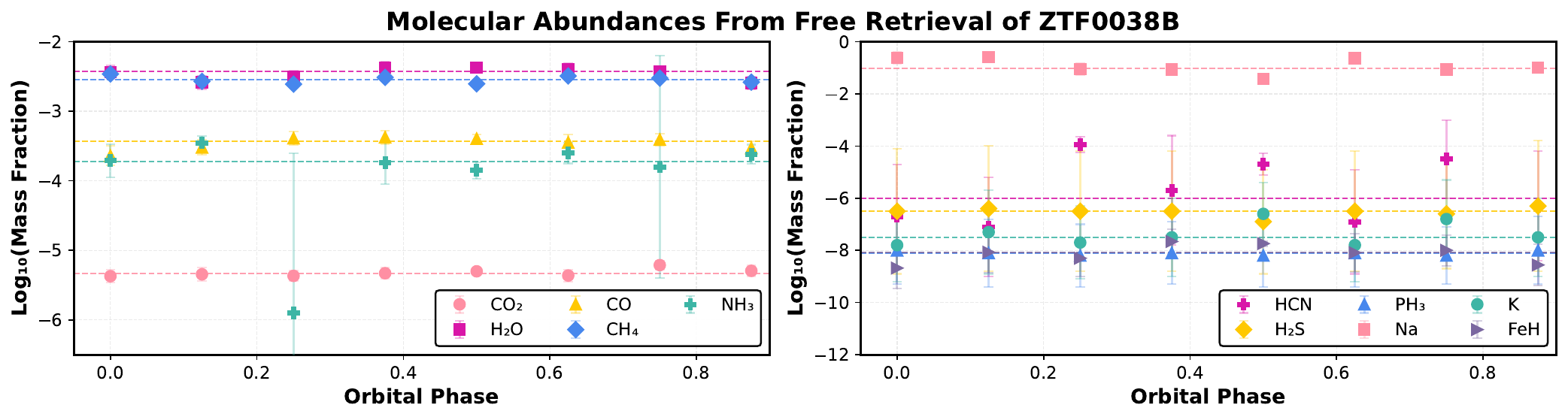}
    \caption{Retrieved abundances for 11 species as a function of phase, as determined via free retrievals for each of the eight phases. Dashed lines indicate the median abundances for each species.}
    \label{fig:free_retrieval_abundances}
\end{figure*}

Because the joint, chemically-constrained retrievals were unable to fit CO$_2$, we also performed free retrievals, where we included vertically-constant mass fractions of H$_2$O, CH$_4$, CO, CO$_2$, NH$_3$, PH$_3$, H$_2$S, HCN, NA, K, and FeH as free parameters. Because of the number of parameters added, we ran each phase individually rather than jointly together, however we fixed the surface gravity, radius, and internal temperature to the best-fit values from the joint retrieval.

Figure~\ref{fig:model_fits} shows the best-fit spectra from the individual-phase free retrieval, demonstrating that the free retrievals can fit the CO$_2$ wavelength range unlike the chemically-constrained retrievals and forward models. Figure~\ref{fig:free_retrieval_abundances} shows the retrieved abundances as a function of longitude. We note that the free retrieval yields nonphysically large abundances for Na. H$_2$O and CH$_4$ are clearly the main atmospheric species, with both varying only modestly across the phases. The CO/CH$_4$ ratio appears stable as a function of longitude at a value of about 0.1. We retrieve a log$_{10}$ CO$_2$ mass fraction of $-5.4 \pm 0.15$ (or log$_{10}$ volume mixing ratio of $-6.6 \pm 0.15$) that also appears to be relatively constant with longitude.

If we add up the number of oxygen and carbon atoms in H$_2$O, CO, CO$_2$, and CH$_4$ implied by the free retrieval, we get results consistent with the chemically-constrained retrieval. Using \texttt{ExoComp} \citep{Lothringer:2026} with the nightside abundances, we measure a $[M/H] = -0.033 \pm 0.012$ and C/O$= 0.67 \pm 0.05$. If we account for 22\% of the oxygen being sequestered into condensates (e.g., silicate clouds) similar to \cite{Line:2021}, we measure $[M/H] = 0.05 \pm 0.012$ and C/O$= 0.53 \pm 0.05$, almost precisely solar abundances.

We also note that the technique of fitting chemical equilibrium solutions to freely retrieved abundances as in \cite{Lothringer:2026} works poorly in this regime due to the effects of chemical disequilibrium, which is presumably the cause of the observed abundance of the carbon species. Nonetheless, our freely retrieved $\log_{10}$ H$_2$O abundance of $-3.26 \pm 0.03$ is in agreement with the expected abundance at [M/H] = 0.0 and C/O = 0.54 (solar) at $\log_{10, VMR}={-3.18}$. The agreement improves further when accounting for 22\% oxygen sequestration in silicates, which brings the expected H$_2$O abundance to $\log_{10, VMR}={-3.30}$.

\section{\SI{4.2}{\micro\meter} Feature Asymmetry}
\label{sec:CO2_asymm}

\begin{figure*}
    \centering
    \includegraphics[width=1\linewidth]{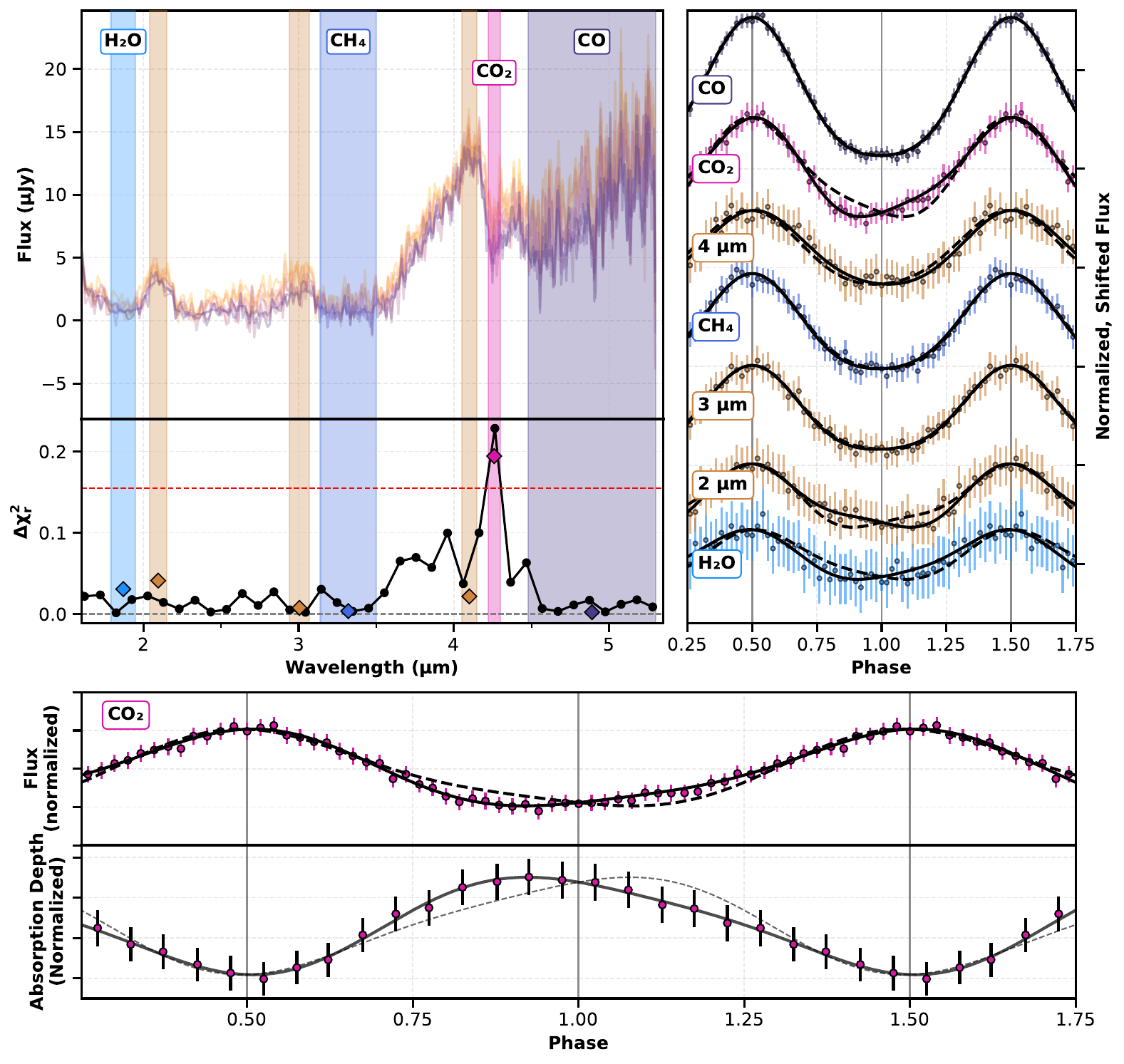}
    \caption{An overview of the observational detections of an asymmetric CO$_2$ absorption feature. The top left panel shows the phase resolved absorption spectra from 1.6 - 5.3 $\upmu$m with highlighted regions corresponding to molecular absorption features and continuum regions. The phase curves for these regions are plotted in the upper right panel. Each phase curve is shown with the best-fitting phase curve model (as discussed in Sections \ref{subsec:broadband_pc_analysis} and \ref{subsec:spec_pc_analysis}) as well the model reflected across the phase = 0.5 (the substellar point). The middle left panel shows the difference in the chi squared value of the original and reflected model fit. The red dashed line represents the threshold for a significant detection. Only the 4.2 $\upmu$m CO$_2$ feature exceeds this threshold. The lower panels show the light curve of the CO$_2$ feature (above) and the depth of the absorption feature as a function of phase (below). Both panels show a western nightside offset that occurs $\sim50$ minutes before the antistellar point}
    \label{fig:CO2_figure}
\end{figure*}

We observe asymmetric phase curves centered around 4.2~$\upmu$m, which coincides with a strong CO$_2$ absorption feature. This asymmetry is visible in Figure \ref{fig:spec_pcs}, and it is detected in the phase offsets at 4.2 $\upmu$m in Figure \ref{fig:amplitude_phase_vs_wavelength}. We quantify the significance of the asymmetry by comparing the reduced chi-squared statistic of the original Fourier model  ($\chi_\text{r, Fourier}^2$) and a reflection of the Fourier model centered on phase = 0.5 ($\chi_\text{r, reflected}^2$). This phase corresponds to where the substellar point is in the direct line-of-sight of the observation (phase E shown in Figure \ref{fig:phase_illustration}). The difference between these statistics ($\Delta\chi^2_\text{r}$) describes how well (or poorly) a reflected Fourier model describes the observed phase curves:

\begin{equation}
    \Delta \chi^2_\text{r} = \chi_\text{r, reflected}^2 - \chi_\text{r, Fourier}^2 
\end{equation}

For our statistical test, the degrees of freedom (dof) are determined by the number of data points in the light curve $(n =820$), and the number of parameters in the Fourier series model ($k = 5$):
\begin{equation}
    \text{dof} = n-k = 815
\end{equation}
For a tidally locked atmosphere with inefficient heat transport, the expected temperature structure would yield a symmetric light curve with a minimum during the mid-eclipse time (phase =0) and a maximum at the substellar point (phase = 0.5). If a light curve is indeed symmetric, then $\chi_\text{r, reflected}^2$ should be $\sim1$. We use this fact to test whether $\Delta \chi^2_\text{r}$ is significant for a given wavelength bin. We use a statistical test assuming the null hypothesis that the data is symmetric (so $\Delta \chi^2_\text{r}$ should be statistically insignificant). A 3$\sigma$ deviation from this assumption (for 815 degrees of freedom) is associated with $\Delta \chi^2_\text{r} > 0.155$. 

We utilize this analysis for all of the spectroscopic phase curves beyond 1.7 $\upmu$m. Because we implement the uncertainty rescaling discussed in Section \ref{subsec:lc_production}, $\chi^2_\text{red, Fourier} \equiv1$. This ensures that $\Delta \chi^2_\text{r}$ is not biased by wavelength dependent systematics that are not accounted for in the pipeline-produced uncertainties. We only detect statistically significant asymmetry in the wavelength bin at 4.2 $\upmu$m, corresponding with the sharp CO$_2$ absorption feature. 

In addition to the evenly-binned spectroscopic phase curves, we use the same asymmetry test to investigate wavelength bands that correspond to pseudo-continuum and absorption features in the spectrum. These regions are highlighted in Figure \ref{fig:CO2_figure}. The only feature with significantly detected asymmetry is the CO$_2$ absorption feature. The minimum point of the Fourier series model of this light curve is located at a phase of 0.917, which occurs 51.6 minutes before the antistellar point.

In addition to this phase curve asymmetry, we also find that the absorption depth of the CO$_2$ feature has asymmetric modulation. For this analysis, we divide the phase into 20 bins and median-combine all integrations within a given phase bin (using Equation \ref{eqn:median_combined_err} to calculate the per-pixel uncertainties). Each of these median-combined spectra contains $\sim 40$ integrations. 

We calculate the absorption depth by calculating the difference between the mean flux from the pseudo-continuum at the top of the absorption and the region centered at the bottom of the sharp absorption feature (4.05--4.15 $\upmu$m and 4.21--4.30 $\upmu$m respectively, highlighted in Figure \ref{fig:CO2_figure}). In order to isolate the behavior of the absorption feature (as opposed to the general photometric modulation), we normalize these median-combined spectra to the pseudo-continuum region at the top of the absorption feature. The uncertainty of the mean flux in each wavelength range is calculated by adding the per-pixel uncertainties in quadrature and dividing by the total number of wavelength elements included in the calculation. From this, the uncertainty of the absorption depth is calculated as the uncertainty from both wavelength windows added in quadrature. 

As a function of phase, this feature exhibits asymmetric behavior similar to the spectroscopic light curve at the CO$_2$ absorption feature. The data is well-fit with a second order Fourier series, and shows clear asymmetry around phase = 0, the nightside phase. The maximum absorption depth is located at a phase of 0.922, which occurs 48.5 minutes before the antistellar point,  further strengthening the detection of asymmetric behavior on the nightside of ZTF0038B.

\section{Energy Balance }\label{sec:energy_balance}

\begin{figure*}[!ht]
    \centering
    \includegraphics[width=0.9\linewidth]{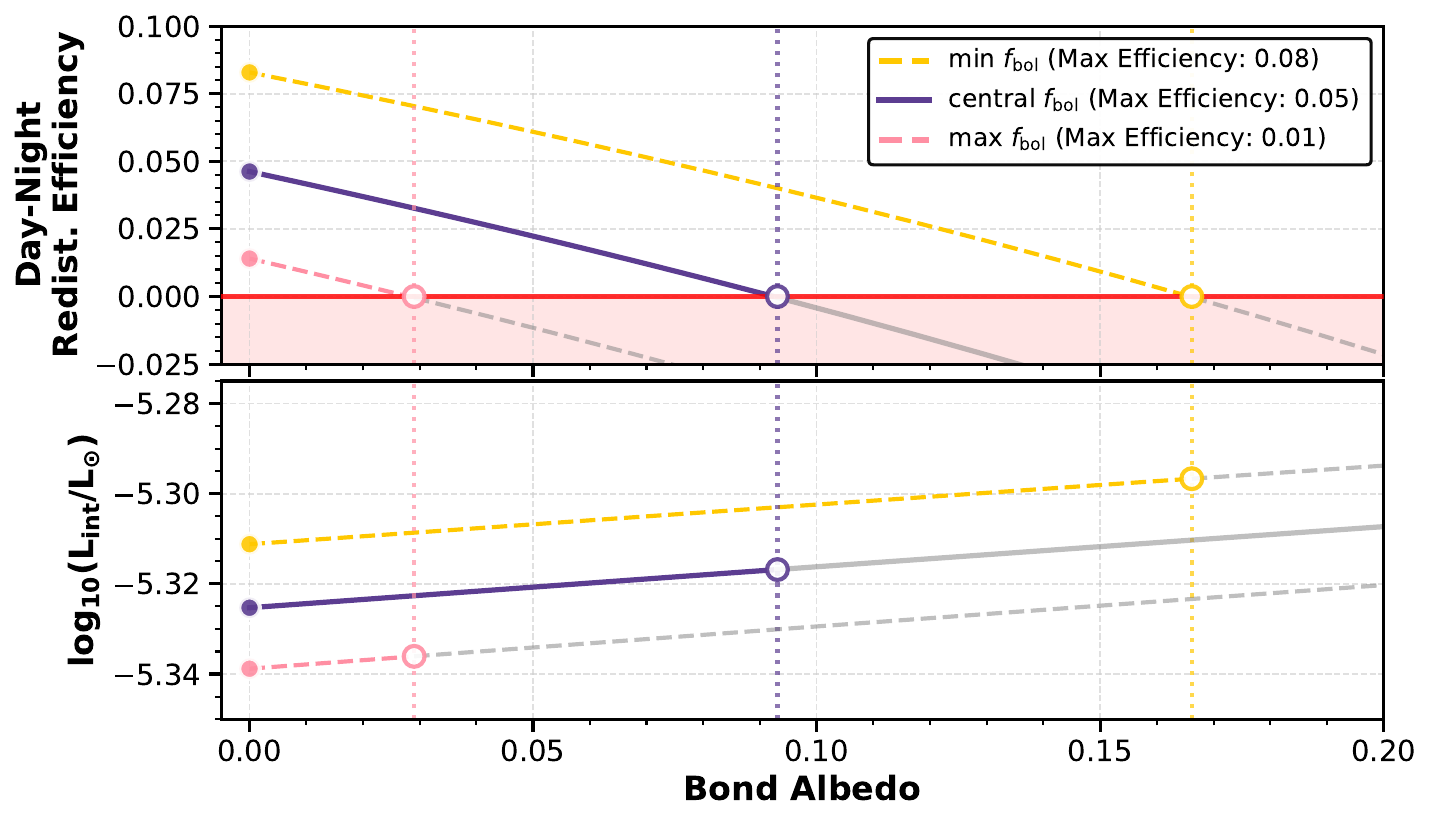}
    \caption{Constraints on bond albedo, day-night redistribution efficiency, and internal luminosity from our energy balance calculations. In both panels, three calculations are shown using a range of plausible bolometric corrections ($f_\text{bol}$). The maximum and minimum $f_\text{bol}$ represent the nominal uncertainty range for the central, fiducial calculation. \textbf{Top:} Day-night redistribution efficiency represents the fraction of the irradiation energy that was absorbed on the dayside and redistributed to the nightside. Negative efficiencies are unphysical; maximum albedo occurs where efficiency reaches zero (open circles, vertical dotted lines). \textbf{Bottom:} Internal luminosity is shown as a function of albedo. For each scenario, the luminosity in the lower panel is constrained by the albedo cut-off from the top panel. }
    \label{fig:L_int_and_evolution}
\end{figure*}

\begin{deluxetable}{lccccccccc}
\tablecaption{Results of the energy balance calculations for three bolometric correction scenarios: using the central bolometric coverage fraction (central $f_\text{bol}$), as well as the maximum and minimum bolometric coverage fractions adopted from the nominal uncertainties discussed in Section \ref{sec:energy_balance}. The effective temperatures of the dayside and nightside hemisphere ($T_\text{eff}$), as well as the temperature contrast ($\Delta T_\text{eff}$), are listed for each bolometric correction scenario. The constraints on the bond albedo ($A_B$), day-night redistribution efficiency ($\eta_\text{dn}$), and interior luminosity ($L_\text{int}$) come from Figure \ref{fig:L_int_and_evolution}.
\label{tab:energy_balance}}
\tablewidth{0pt}
\tablehead{
    \colhead{} & 
    \colhead{$f_\mathrm{day}$} &
    \colhead{$f_\mathrm{night}$} &
    \colhead{$T_\mathrm{day}$ (K)} &
    \colhead{$T_\mathrm{night}$ (K)} &
    \colhead{$\Delta T_\mathrm{eff}$ (K)} &
    \colhead{$A_B$} & 
    \colhead{$\eta_\text{dn}$} & 
    \colhead{$\log_{10}(L_\mathrm{int}/L_\odot)$} 
}
\startdata
Min $f_\text{bol}$   & 0.788 & 0.806 & 1052 & 980 & 72 & $< 0.17$ & $< 0.08 $& $[-5.311,\ {-5.297}]$ \\
Central $f_\text{bol}$  & 0.798 & 0.844 & 1049 & 968 & 81 & $ < 0.09$ & $<0.05$ & $[-5.325,\ {-5.317}]$ \\
Max $f_\text{bol}$   & 0.808 & 0.882 & 1046 & 958 & 88 & $< 0.03$ & $<0.01$ & $[-5.339,\ {-5.336}]$ \\
\enddata
\end{deluxetable}

We use our best-fit free retrieval models of the dayside and nightside to calculate bolometric corrections for our NIRSpec PRISM observations. We extend the retrieved spectra from 0 - 20 $\upmu$m to cover the full spectral energy output from the brown dwarf, and we calculate how much of the bolometric flux of the brown dwarf is covered by our \textit{JWST} observations. For the calculation of the bolometric flux, we clip the wavelength coverage of our observations to be between 0.95--5.0 $\upmu$m. This excludes the noisy short wavelengths that don't cover substantial brown dwarf contribution, as well as longer wavelengths where the unexplained 5 $\upmu$m flux deviates substantially from all models. The bolometric corrections for the dayside and nightside are calculated as the fraction of the integrated brown dwarf emission that is covered between 0.95--5.0 $\upmu$m. We refer to this quantity as the bolometric coverage fraction, $f_\text{bol}$. We find that 79.8\% of the dayside flux and 84.4\% of the nightside flux are covered in our observation ($f_\text{bol, day}$ = 0.798 and $f_\text{bol, night}$ = 0.844 ).

We use these bolometric coverage fractions to calculate dayside and nightside effective temperatures using the Stefan-Boltzmann law for hemispherical observations:
\begin{equation}
    \sigma T_\text{eff}^4= 2  \cdot (\frac{d_\text{BD}}{ R_\text{BD}})^2\cdot \frac{1}{f_\text{bol}}\int F_\nu d\nu
\end{equation}
The effective dayside and nightside temperatures are $T_\text{eff, day} = 1049$ K and $T_\text{eff, night} =968$ K, respectively.

We adopt nominal uncertainty estimates by determining how the bolometric correction of a blackbody would change given an assumed temperature uncertainty. We use the effective temperature of the dayside and nightside hemispheres and adopt the temperature uncertainty from our retrieved temperature-pressure profiles at a pressure level of 1 bar (as shown in Figure \ref{fig:retrieval_fig}). The nightside uncertainty at 1 bar is $20$ K, and the dayside uncertainty at 1 bar is $6$ K. The brightness temperature of the dayside and nightside spectrum indicates that our observations probe pressures on the order of 1-10 bars (Figure \ref{fig:brightness_temp}), and the temperature uncertainty is largest at lower pressures (Figure \ref{fig:retrieval_fig}). For this reason, our choice to adopt the temperature uncertainty at 1 bar is a conservative choice, and even so, the temperature profile remains quite well-constrained. We use our effective dayside and nightside temperatures and nominal temperature uncertainties ($T_\text{night} =968 \pm20$ K and $T_\text{day} = 1049 \pm6$ K) to determine a plausible uncertainty range for our bolometric corrections. Using these temperatures and uncertainty estimates, the associated blackbody would have an uncertainty of $\sigma_{f_\text{bol, day}} = 0.0033$ for the dayside hemisphere, and $\sigma_{f_\text{bol, night}} = 0.0128$ for the nightside. 

Using these nominal uncertainties, we carry out the subsequent energy balance calculations for three cases: 
\begin{enumerate}
    \item Central $f_\text{bol}$: Adopt the fiducial $f_\text{bol}$ values for both hemispheres as calculated from the best-fit free retrieval models for the dayside and nightside. $f_\text{central, day}$ = 0.798 and $f_\text{central, night}$ = 0.844. 
    \item Min $f_\text{bol}$: Adopt the minimum $f_\text{bol}$ values assuming $3\sigma$ uncertainties ($f_\text{min} =f_\text{central} - 3\sigma_{f} $ for the dayside and nightside). $f_\text{min, day}$ = 0.788 and $f_\text{min, night}$ = 0.806.
    \item Max $f_\text{bol}$: Adopt the maximum $f_\text{bol}$ values assuming $3\sigma$ uncertainties (($f_\text{max} =f_\text{central} +3\sigma_{f} $ for the dayside and nightside). $f_\text{max, day}$ = 0.808 and $f_\text{max, night}$ = 0.882.
\end{enumerate}

We find the bolometric flux for the dayside and nightside are $F_\text{bol, day}= 1.10\times 10^{-14}$ erg/s/cm$^2$, and  $F_\text{bol, night}= 7.98 \times 10^{-15}$ erg/s/cm$^2$. The bolometric luminosity for the dayside and nightside hemispheres is calculated as:

\begin{equation}
    L_\text{hemisphere} = 4 \pi \cdot d^2 \cdot \frac{1}{f_\text{bol}}\int F_\nu d\nu
\end{equation}

We find the dayside and nightside luminosities to be $L_\text{bol, day} = 1.27 \times 10^{28}$ erg/s and $L_\text{bol, night} = 9.23 \times 10^{27}$ erg/s. The total bolometric luminosity is the sum of the luminosity of both hemispheres: $L_\text{bol} = 2.19 \times 10^{28}$ erg/s ($\log_{10}(L_\text{bol}/L_\odot)=-5.24$) . Using the measured bolometric luminosities of the brown dwarf's dayside and nightside, we constrain several key components of its energy budget. First, the interior luminosity ($L_\text{int}$) represents the brown dwarf's internal heat, independent of irradiation from the white dwarf.  We calculate $L_\text{int}$
as the total observed emission from both hemispheres minus the absorbed stellar energy:
\begin{equation} \label{eqn:Lint}
    L_\text{int} =  L_\text{bol, day} + L_\text{bol, night} - L_\text{irr}.
\end{equation}

The irradiation luminosity ($L_\text{irr}$) corresponds to   
the stellar energy absorbed by the brown dwarf from the white dwarf. This depends on the incident flux ($F_\text{incident}$), the radius of the brown dwarf ($R_\text{BD}$), and the Bond albedo ($A_B$):
\begin{equation} \label{eqn:Lirr}
    L_\text{irr} = (1-A_B) \cdot F_\text{incident} \cdot \pi R_\text{BD}^2.
\end{equation}

Some of $L_\text{irr}$ is circulated from the dayside to the nightside. We define this transported energy as $L_\text{cir}$,  
which we calculate as the nightside luminosity minus half the internal luminosity. Based on the energy balance of the nightside, $L_\text{cir}$ can be expressed as: 
\begin{equation}\label{Lcir}
    L_\text{cir} = L_\text{night} - \frac{L_\text{int}}{2}.
\end{equation}
Half of $L_\text{int}$ is subtracted because only half of the internal luminosity is emitted from the nightside hemisphere. We define the day-night redistribution efficiency ($\eta_\text{dn}$), which quantifies the fraction of the incident energy from the white dwarf that is redistributed and re-emitted on the nightside:
\begin{equation} \label{eqn:efficiency}
    \eta_\text{dn} = \frac{L_\text{cir}}{L_\text{irr}} = \frac{L_\text{night} - \frac{1}{2}L_\text{int}}{(1-A_B)\cdot F_\text{incident}\cdot \pi R^2_\text{BD}}  .
\end{equation}

These calculations constrain the internal luminosity and redistribution efficiency as a function of albedo. We calculate these values for three scenarios using the three bolometric coverage assumptions (minimum $f_\text{bol}$, central $f_\text{bol}$, and maximum $f_\text{bol}$). Figure \ref{fig:L_int_and_evolution} and Table \ref{tab:energy_balance} present the calculation results. This calculation suggests that the brown dwarf has a low bond albedo ($A_B \lesssim 0.2$), low day-night redistribution efficiency ($\eta_\text{dn} \lesssim 0.08$), and a well constrained interior luminosity ($\log_{10}{(L_\text{int}/L_\odot}) =-5.32$).

\section{Discussion}\label{sec:discussion}

\subsection{Disequilibrium Processes Homogenize CO and CO$_2$ Abundances}

\begin{figure*}
    \centering
    \includegraphics[width=0.85\linewidth]{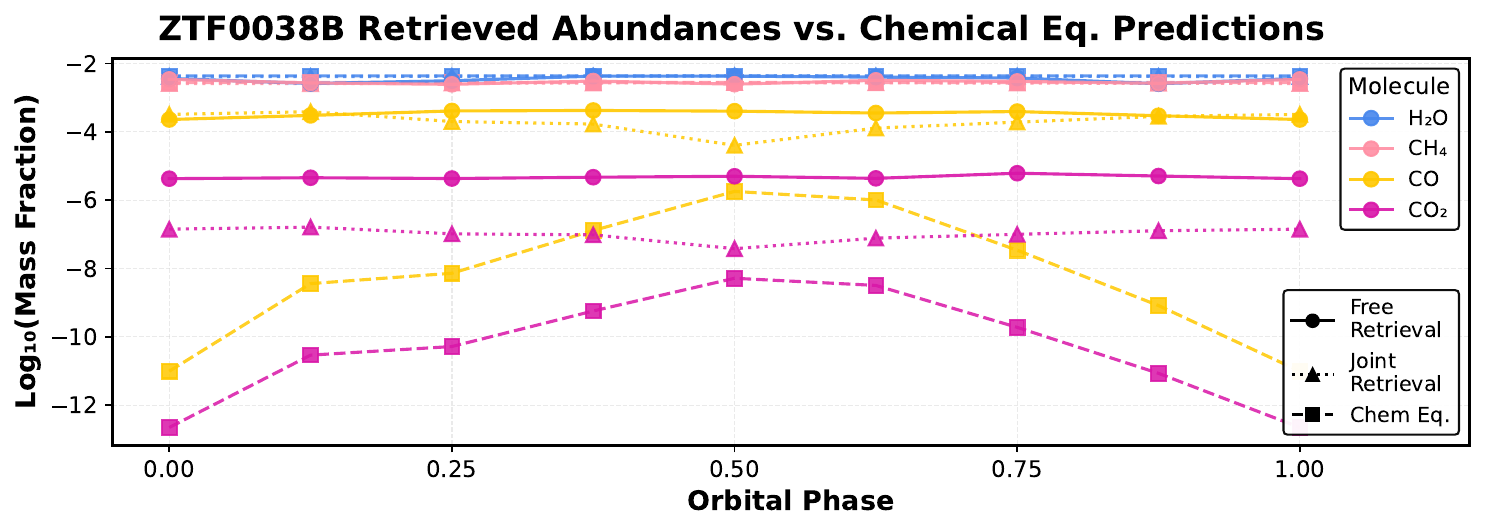}
    \caption{Comparison of the retrieved abundances and predictions from chemical equilibrium. H$_2$O (blue), CH$_4$ (pink), CO (yellow), and CO$_2$ (fuchsia) abundances are shown from the free retrieval (circles), joint retrievals (triangles), and the chemical equilibrium prediction (squares).  H$_2$O and CH$_4$ are both consistent with the chemical equilibrium predictions, while the retrieved abundances for both CO and CO$_2$ are larger and more uniform than the chemical equilibrium prediction. The CO$_2$ abundances from the joint retrieval should be interpreted with caution, as the joint retrieval struggles to fit the 4.2 $\upmu$m CO$_2$ absorption feature.}
    \label{fig:comp_abundances_cheq}
\end{figure*}
Figure \ref{fig:comp_abundances_cheq} compares the longitudinally-resolved abundances from the chemically-constrained and free retrievals, as well as chemical equilibrium predictions. We obtain the abundances in the chemically-constrained retrievals by calculating the abundance at the temperature and pressure of the quench level at the retrieved metallicity and C/O ratio. We compare these to the abundance in chemical equilibrium (i.e., without a quench pressure) by calculating the abundances at the retrieved 1 bar temperature with the retrieved metallicity and C/O ratio.
 
Among retrieved molecules, CO and CO$_2$ demonstrate evidence for disequilibrium chemistry. Assuming pure chemical equilibrium, we would expect the abundances of CO and CO$_2$ to vary substantially between the dayside and nightside due to different temperature-pressure profiles between the two sides (Figure \ref{fig:comp_abundances_cheq}). However, the retrieved abundances using our free retrieval, which provides the best fit to the CO$_2$ feature, indicate that the key molecular species are relatively constant across the atmosphere. Such deviations from chemical equilibrium predictions are detected in JWST observations of hot Jupiters, where they are attributed to vertical and horizontal transport quenching the chemical processes \citep{bell_nightside_2024, parmentier_horizontal_2026}. ZTF0038B occupies a higher-gravity, faster-rotating regime than previously studied systems. Our phase-resolved spectra also enable us to probe the relative contributions of vertical and horizontal transport at higher S/N than previously possible.

Our joint atmospheric retrieval assumes chemical equilibrium at altitude below the quench layer and uniform abundances above it. The quench pressure is a free parameter fit independently at each of the 8 phases. For ZTF0038B, it is $\sim$7-30 bars. The abundances of CO from our joint retrieval are orders-of-magnitude larger than the chemical equilibrium abundance. CO is more abundant deeper in the atmosphere. When the mixing ratios are frozen at $P=P_\text{quench}$, the retrieved CO abundances therefore exceed the chemical equilibrium prediction. Our joint retrieval configuration homogenizes the CO abundances by varying the quench pressure accordingly. However, in the hot Jupiter regime, horizontal winds are the physical process that is expected to be responsible for homogenization from the dayside to the nightside \citep{parmentier_horizontal_2026}. In our analysis, we cannot disentangle which physical process is responsible for the observed abundances, but the homogeneous abundance of CO is robustly detected in the free retrieval and joint retrieval.

The CO$_2$ feature at 4.2 \micron{} is not well-fit in the joint retrieval. We therefore turn to the free retrieval to assess CO$_2$. The free retrieval analysis assumes vertically-constant abundances at each phase, provides an improved fit, and supports a similar interpretation for CO and CO$_2$. We find that CO and CO$_2$ abundances are both larger and more homogenized than the prediction from chemical equilibrium models (Figure \ref{fig:comp_abundances_cheq}).

\subsection{Anomalous Features in the Phase Curve Observations of ZTF0038B}

We fit the brown dwarf's emission spectra with forward model grids, a joint-phase atmospheric retrieval, and compare the nightside with NIRSpec PRISM observations of non-irradiated brown dwarf atmospheres. In all of these analyses, we find that ZTF0038B is relatively well-fit at shorter wavelengths, but deviates substantially from grid and retrieval models beyond $\sim$4 $\upmu$m (see Figures \ref{fig:model_fits} and \ref{fig:retrieval_fig}). This wavelength range has three dominant features: a CO$_2$ absorption feature around 4.2 $\upmu$m, a CO absorption feature around 4.5 $\upmu$m, and a broad, unidentified excess flux around 5 $\upmu$m. It is unclear whether this excess is due to systematics at the edge of the detector. However, this wavelength range coincides with the H$_3^+$ fluorescence band. External radiation drives similar emission features in solar system planets \citep{giles_detection_2016, drossart_detection_1989, oka_observations_1990, trafton_detection_1993, nichols_dynamic_2025, melin_discovery_2025}. If verified, this would provide the first compelling evidence for external radiation-driven emission associated with non-equilibrium ion chemistry outside of our solar system—a discovery that would fundamentally expand our understanding of atmospheric energy balance in irradiated substellar objects.

The CO$_2$ absorption feature is discrepant with most atmospheric models and template comparisons for ZTF0038B (Figures~\ref{fig:field_bd_comparison} and \ref{fig:model_fits}). The CO$_2$ light curve and time-series absorption feature reveal a significant deviation from the general atmospheric behavior in all other observed wavelengths (Figure~\ref{fig:CO2_figure}). In Section \ref{sec:CO2_asymm}, we find that the spectroscopic light curves centered around this absorption feature are asymmetric; the dayside peak is centered at the substellar point, but there is a significant Eastern shift in the phase curve's point of minimum flux. We find a similar nightside asymmetry in the normalized depth of the absorption feature as well (discussed in Section \ref{sec:CO2_asymm}). Both of these phase-resolved analyses reveal a nightside offset of $\sim50$ minutes prior to the antistellar point. 

This asymmetric light curve structure at 4.2 $\upmu$m is surprising. Previous general circulation models (GCM) coupled with a mini chemical scheme predict relatively uniform nightside CO$_2$ distribution in two hot Jupiters \citep{lee_mini-chemical_2023}. The longitudinal abundances of CO$_2$ produced by the free retrieval indicate relatively constant column-integrated abundances (Figure~\ref{fig:free_retrieval_abundances}), which suggests that the observed asymmetry is more likely due to a temperature asymmetry. However, light curve features associated with a temperature asymmetry should be apparent at additional wavelengths other than just this 4.2 $\upmu$m absorption feature. We do not observe this asymmetric feature in the spectroscopic phase curves at any other wavelengths (Figure \ref{fig:CO2_figure}).

The light curve asymmetry, if caused by a non-uniform distribution of CO$_2$, might be attributed to the fact that CO$_2$ is highly sensitive to the vertical mixing strength \citep{Mukherjee2024, beiler_precise_2024}. However, this explanation is difficult to reconcile with our findings that the retrieved column-integrated CO$_2$ abundances are nearly constant from the dayside to the nightside. Understanding the physical origin of this asymmetry likely requires disentangling the impact of three-dimensional thermal structure and chemical composition on the disk-integrated spectrum, a capability beyond the scope of our retrieval analysis. This surprising asymmetric light curve shape, unveiled by our high S/N phase curves, calls for further modeling of disequilibrium processes to determine the physical mechanism behind this finding.

\subsection{Inefficient Day-Nightside Heat Transport}\label{subsec:circulation}
ZTF0038B shows highly inefficient day-to-nightside heat transport, supported by four independent lines of evidence. The spectral features of the nightside are fully consistent with those of field brown dwarfs of similar spectral type (Figure~\ref{fig:field_bd_comparison}), indicating that day-to-night circulation is insufficient to produce detectable departures from a non-irradiated atmosphere.

The phase curve shape provides a second independent line of evidence for inefficient heat transport. From the broadband and spectroscopic phase curve analyses, we find that the preferred phase curve model is a second order Fourier series in which the phase of the $k=2$ component peaks at the substellar and antistellar points. This is opposite to the geometric distortion expected from tidal deformation, which would flatten the dayside peak and deepen the nightside trough \citep{Morris1985,Shporer2019}.  When heat transport is inefficient, absorbed irradiation remains localized and re-radiates near the substellar point rather than being homogenized across the dayside. This produces a steep brightness gradient on the dayside and a relatively uniform nightside, which is precisely described by the $k=2$ term \citep{zhou_hstwfc3_2021}. Visually, this is manifested in the phase curve as a sharp peak and a flattened trough, as shown in Figures \ref{fig:spec_pcs} and \ref{fig:broadband_phase_curve}. The phase curve shape therefore indicates that hemispheric temperature distributions are governed primarily by local radiative heating and cooling rather than horizontal transport.

The absence of a hotspot offset provides a third independent line of evidence for inefficient heat transport. The spectroscopic phase curve peaks align with the substellar point within $<2\sigma$ between 1.7 - 5.3~$\upmu$m, where significant brown dwarf day-night modulation is detected. The phase curve peaks show no hotspot offset at any wavelength, indicating the absence of a strong zonal jet capable of producing the eastward temperature shifts widely observed in slowly rotating hot Jupiters \citep{Knutson2007,Stevenson2014,mikal-evans_jwst_2023,bell_nightside_2024}. This is consistent with the expectation from GCMs due to ZTF0038B's Jupiter/Saturn-like rotation period. At this rotation rate, the strong Coriolis force confines the pressure gradient balance to progressively lower latitudes, yielding a small equatorial deformation radius ($L_D \propto \Omega^{-1/2}$) and a correspondingly narrow equatorial jet with minimal longitudinal heat transport. \citep{Parmentier2018haex.bookE.116P,TanShowman2020ApJ...902...27T,Lee2020}. This regime differs fundamentally from slowly rotating hot Jupiters, where broad zonal flows redistribute a substantial fraction of absorbed stellar energy and produce the eastward hotspot offsets \citep{Showman2009ApJ...699..564S, Rauscher2010ApJ...714.1334R, Heng2011MNRAS.413.2380H, Showman2011ApJ...738...71S, Mayne2014A&A...561A...1M}.

The fourth and most direct line of evidence comes from the energy balance calculation enabled by the exceptional precision and broad wavelength coverage of the \textit{JWST} spectra. The PRISM coverage captures $\sim80$\% of the brown dwarf's bolometric emission, allowing a nearly model-independent constraint on the day-night redistribution efficiency ($\eta_\text{dn}$). We constrain $\eta_\text{dn}$ to be $<$ 10\%. This confirms that the vast majority of the absorbed stellar energy is re-radiated locally on the dayside rather than transported to the nightside.

\subsection{Precise Constraints on Orbital Evolution, System Age, and Common Envelope Efficiency}\label{subsec:evolution}

\subsubsection{Period Evolution}
We do not detect transit/eclipse timing variations and constrain the orbital period of ZTF0038B to a precision of 60~$\upmu$s, which corresponds to an evolutionary rate of $|\dot{P}/P| \lesssim 1.6 \times10^{-9}$. This precludes the possibility of another massive companion in the system. 

Gravitational wave emission is the only expected source of angular momentum loss in this system  \citep{zorotovic_close_2022}. For a circular orbit, the expected fractional period evolution is \citep{blanchet_gravitational_2014}:
\begin{equation}
\dot{P}/P = -\frac{192\pi}{5c^5} (2 \pi G )^{5/3} \frac{M_\text{WD}M_\text{BD}}{(M_\text{WD}+M_\text{BD})^{1/3}}P^{-8/3},
\end{equation}
with a decay time \citep{peters_gravitational_1964}, defined as the lifetime for the system to collapse: 
\begin{equation}
    T_c (a) = \frac{5c^5a^4}{256G^3 M_\text{WD}M_\text{BD}(M_\text{WD}+M_\text{BD})}.
\end{equation}

For ZTF0038, these yield $\dot{P}/P = -9\times10^{-20}$\,s$^{-1}$ and $T_c \sim 140$\,Gyr, far below current detection thresholds given its 10.4~hr orbital period. However, the steep $P^{-8/3}$ scaling means that shorter-period WD--BD systems \citep[e.g.,][]{Casewell2018MNRAS.476.1405C} will experience substantially stronger orbital decay, and improved eclipse timing precision with \textit{JWST} may soon enable direct detection of gravitational wave-driven period evolution in these systems.

\subsubsection{A Precise System Age Constraint and the Thermal Evolution of an Irradiated Brown Dwarf}

 \begin{figure*}
     \centering
     \includegraphics[width=1\linewidth]{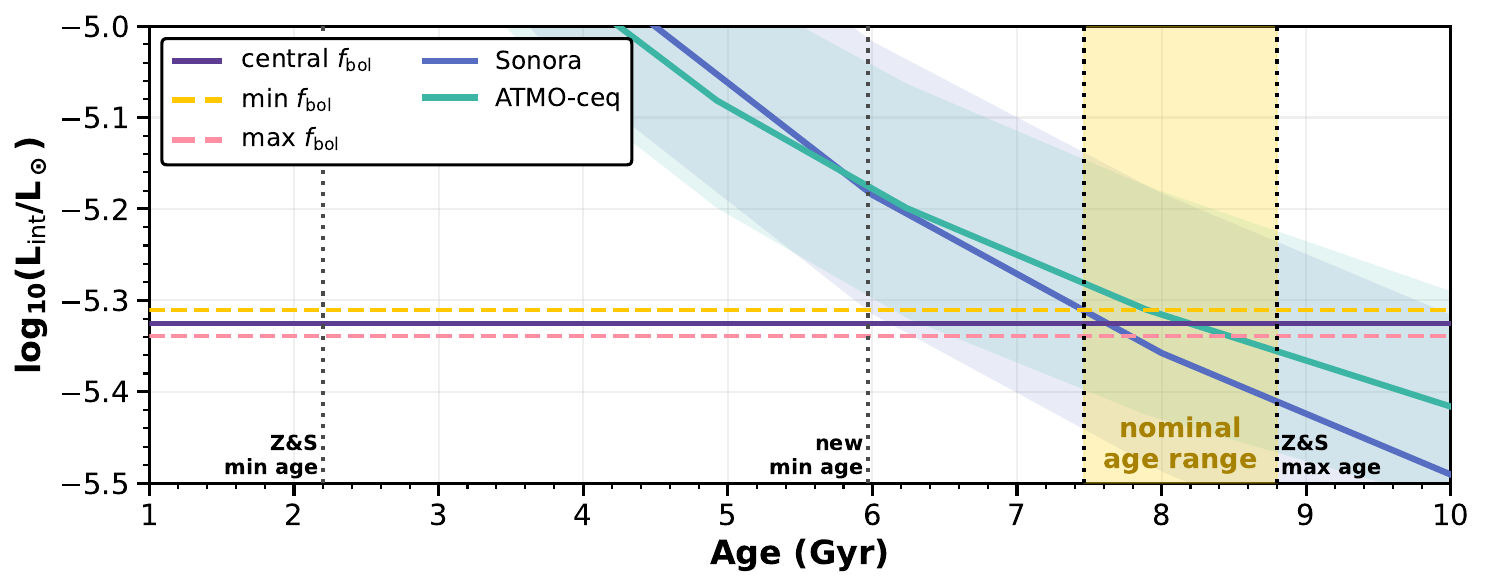}
     \caption{Luminosity constraints from the energy balance calculation for ZTF0038B (horizontal lines) are compared with field brown dwarf evolutionary tracks (solid blue and green curves). The shaded regions around the evolutionary tracks indicate the uncertainty in the evolutionary models based on the mass uncertainty from \cite{van_roestel_ztfj00382030_2021}. The horizontal dashed lines indicate age constraints from this work and \cite{zorotovic_close_2022}. Assuming the mass presented in \cite{van_roestel_ztfj00382030_2021}, we provide a nominal age estimate highlighted in yellow.}
     \label{fig:evolution}
 \end{figure*}
Isolated white dwarf progenitor mass can be predicted by using initial-to-final mass relations and the progenitor mass then informs its main sequence life time \citep{Weidemann2000A&A...363..647W,Cummings2018AJ....156..165C}. However, the presence of a companion during the common envelope phase drives envelope ejection before accretion onto the white dwarf core is complete. This breaks the initial-to-final mass relation \citep{Ivanova2013A&ARv..21...59I}. \citet{zorotovic_close_2022} reconstruct the evolutionary history of ZTF0038 by exploring the full range of viable white dwarf progenitor masses, constraining them to $1.08$--$1.59$~M$_\odot$. The corresponding main sequence lifetimes span $\sim$2.2--8.8~Gyr. They also derive a tentative lower age limit from the brown dwarf's mass and radius. However, this age estimate relies on brown dwarf evolution models, and the radius dependence on age is fairly weak \citep{Burrows2011}.

The bolometric luminosity measured from our \textit{JWST} observations provides a substantially stronger age constraint. Figure~\ref{fig:evolution} compares the observed internal luminosity with evolutionary 
tracks of an isolated $62~$M$_\mathrm{Jup}$ brown dwarf, adopting both the 
Sonora Bobcat and ATMO 2020 grids \citep{Marley2021, phillips_new_2020}  at solar metallicity, consistent with our retrieval results (Sections~\ref{subsec:joint_retrieval} and \ref{subsec:free_retrievals}). 

The common envelope phase may have deposited additional heat into the brown dwarf, delaying its cooling and causing it to appear younger than it truly is. The luminosity therefore yields a lower limit rather than a precise age. Assuming no such heat was deposited gives the most conservative estimate of 7.5~Gyr. Combined with the upper limit from \citet{zorotovic_close_2022}, the predicted age of ZTF0038 is between 7.5--$8.8$~billion years old. However, we note that the uncertainty in the mass estimate from radial velocity observations presented in \cite{van_roestel_ztfj00382030_2021} limit the precision to which we can constrain the age. By accounting for the uncertainty in the evolutionary cooling track due to the mass uncertainty of $\pm 4$ M$_\text{Jup}$ (shown as the shaded regions around both evolutionary models in Figure \ref{fig:evolution}), we set a lower age constraint of 6 Gyr. The age constraint is now limited by the mass estimate of the brown dwarf, which may be improved with continuous radial velocity monitoring.

This age constraint is the most precise age determination for any WD--BD system to date. The close agreement between the observed luminosity and the isolated brown dwarf evolutionary tracks (Figure~\ref{fig:evolution}) further implies that the common envelope phase did not significantly alter the brown dwarf's subsequent thermal evolution. Any heat deposited during the common envelope phase — whether through tidal dissipation, frictional drag, or irradiation — appears negligible compared to the primordial heat reservoir retained from formation. The brown dwarf therefore cools as though it were an isolated field object, despite its dynamically violent evolutionary history.

\subsubsection{Common Envelope Efficiency Constraint}
The common envelope efficiency parameter, $\alpha$, represents the fraction of orbital energy used to eject the envelope. \citet{zorotovic_close_2022} reconstruct the evolutionary histories of a sample of WD--BD binaries to constrain $\alpha$, but are only able to place tentative constraints on individual systems using relatively weak mass-radius-age relations. Our precise age constraint for ZTF0038 breaks this degeneracy. The observed luminosity places the system within the age range explored by \citet{zorotovic_close_2022}, and the combined constraints narrow the common envelope efficiency to $\alpha \sim 0.24$--$0.7$. Our nominal age estimate, highlighted in Figure \ref{fig:evolution}, places an even tighter estimate on the common envelope efficiency, narrowing it to  $\alpha \sim 0.24$--$0.5$.

This result is consistent with binary evolution reconstructions of post-common envelope binaries containing both brown dwarf and main sequence companions \citep{zorotovic_post-common-envelope_2010, zorotovic_close_2022}. These studies find that the observed populations can be explained with $\alpha \lesssim 0.5$. While individual systems in these samples have poorly constrained $\alpha$ values, the population-level result is robust. A precise age measurement for ZTF0038 goes further: low common envelope efficiency may be necessary to explain the observed features of this system. Our nominal age estimate indicates that a low common envelope efficiency is favored, but a tighter mass constraint could provide even more definitive constraints.

\section{Conclusions}\label{sec:conclusion}

We present the first \textit{JWST} phase curve observations of a white dwarf–brown dwarf binary, ZTF0038, using NIRSpec/PRISM. The eclipsing geometry, broad wavelength coverage (0.6--5.3 $\upmu$m), and full orbital phase coverage make this dataset a uniquely detailed probe of a tidally locked, irradiated substellar atmosphere. We summarize our findings below: 

\begin{enumerate}
    \item The phase curve covers a full orbit of the binary, including a total eclipse of the white dwarf. The total eclipse enables a model-independent separation between the brown dwarf and white dwarf spectral contributions at all orbital phases. The resulting phase-resolved spectra monitor the brown dwarf's rotational modulation with a broad wavelength coverage (0.6--5.3 $\upmu$m), high SNR, and high temporal resolution. The observation covers $\sim80\%$ of the bolometric emission. We detect strong rotational modulation at wavelengths $>1.7$\,$\upmu$m. This high amplitude is consistent with a dayside-to-nightside temperature gradient in a tidally locked atmosphere.
    \item From the eclipse photometry and phase-resolved spectroscopy, we constrain key atmospheric and orbital parameters of ZTF0038B. 
    \begin{enumerate}
        \item Spectral fitting with forward model grids favors a cloudless atmospheric model for the brown dwarf.
        \item The dayside and nightside bolometric fluxes are: $F_\text{day} = 1.10 \times10^{-14}$ erg/s/cm$^2$ and $F_\text{night} = 7.98 \times10^{-15}$ erg/s/cm$^2$,  corresponding to effective temperatures of 1049 K and 968 K respectively.
        \item Based on our energy balance calculation, the internal luminosity ($\log(L_{BD}/L_\odot)$) is constrained between $-5.34$ and $-5.30$ and the bond albedo is constrained from 0 to 0.2. 
        \item Our eclipse timing analysis provides a precise constraint on the period evolution of ZTF0038. The period (0.43192080512 $\pm6.7 \times 10^{-10}$ days) is constrained to a precision of $\sim60~\upmu$s, and the fractional period evolution is constrained to $|\dot{P}/P| \lesssim 1.6 \times 10^{-9}.$
    \end{enumerate}
    \item Four independent lines of evidence indicate that the brown dwarf ZTF0038B experiences inefficient heat transport from the dayside to the nightside.
    \begin{enumerate}

        \item The broadband and spectroscopic phase curves show no measurable offset in the hotspot peak relative to the substellar point. Phase curve peak offsets are commonly observed in hot Jupiters and attributed to efficient equatorial jets. The absence of an offset in ZTF0038B is consistent with jet suppression driven by its fast rotation.

        \item The broadband and spectroscopic phase curves are well-fit by a second-order Fourier series with a flat trough and a sharper peak. This shape indicates that dayside irradiation remains localized near the substellar point, rather than being redistributed efficiently to the nightside.

        \item The nightside emission of ZTF0038B is consistent with both observations and spectral models of non-irradiated late-T type brown dwarfs. This indicates that neither the dayside irradiation nor the common envelope evolution of ZTF0038B significantly altered the nightside atmosphere of ZTF0038B.
        \item Energy balance calculations constrain the dayside-to-nightside heat transport efficiency of ZTF0038B to $<10\%$, placing it among the least efficient heat transporters observed in tidally locked atmospheres.
    \end{enumerate}
\item Phase-resolved atmospheric retrievals indicate that the CO and CO$_2$ abundances are both larger and more uniform than chemical equilibrium predictions. This suggests that horizontal and/or vertical circulation processes are homogenizing these key species in the atmosphere.

\item The 4.2 $\upmu$m CO$_2$ absorption band phase curve shows a nightside asymmetry. The dayside peak is centered on the substellar point, but the nightside trough is offset from the anti-stellar point. This nightside offset is independently confirmed by both the phase curve centered on the absorption band and the normalized absorption depth as a function of orbital phase. This feature could be attributed to CO$_2$ being sensitive to vertical mixing, but this interpretation is difficult to reconcile with the uniform CO$_2$ abundances obtained from our free retrieval analysis.

\item We observe excess emission at the longest wavelengths covered by PRISM. This feature is not well-fit by forward models or the atmospheric retrievals. The excess coincides with the H$_3^+$ fluorescence band, though an instrumental systematic near the detector edge cannot be ruled out. If the H$_3^+$ interpretation is confirmed, it would represent the first detection of externally driven non-thermal emission in an extrasolar atmosphere.

\item By comparing the observed luminosity of ZTF0038B with isolated brown dwarf evolutionary tracks, we derive a precise binary age of 7.5–8.8 Gyr, as well as a robust lower age limit of 6 Gyr. This constraint, applied to the reconstructed evolutionary history of ZTF0038 \citep{zorotovic_close_2022}, pins the common envelope efficiency to $\alpha_{\mathrm{CE}}\sim$ $0.24-0.7$, one of the tightest observational constraints on this poorly understood phase of binary stellar evolution. 
\end{enumerate}

ZTF0038B is a brown dwarf placed in a very unique and extreme environment. It is a survivor of common envelope evolution, and a recipient of strong irradiation on its tidally locked dayside that resembles warm Jupiters. Even though the dayside receives irradiation from the white dwarf, the atmospheric characteristics of ZTF0038B are remarkably consistent with current frameworks of brown dwarf and post-common envelope binary evolution. Despite its evolutionary history and irradiation environment, the nightside of ZTF0038B behaves much like a non-irradiated brown dwarf, making it a uniquely well-characterized benchmark for post-main sequence binaries and planetary systems. As the first of five WD--BD binaries in an ongoing JWST survey, this study offers an exciting demonstration of what phase curve observations can reveal about how irradiation and common envelope evolution shape substellar atmospheres across a range of system parameters.

\begin{acknowledgements}
We thank the anonymous referee for a constructive and timely report that improves the quality of the manuscript. We thank Maddie Chapleski for creating the initial version of Figure \ref{fig:field_bd_comparison}.
This work is based on observations made with the NASA/ESA/CSA James Webb Space Telescope. The data were obtained from the Mikulski Archive for Space Telescopes at the Space Telescope Science Institute, which is operated by the Association of Universities for Research in Astronomy, Inc., under NASA contract NAS 5-03127 for JWST. The specific observations analyzed can be accessed via \dataset[doi: 10.17909/0z03-1m62]{https://doi.org/10.17909/0z03-1m62}. These observations are associated with program JWST-GO-4967. Support for this work was provided by NASA through a grant from the Space Telescope Science Institute under that program. DBL acknowledges the support from the Virginia Space Grant Consortium through the Graduate STEM Research Fellowship Program.
\end{acknowledgements}

\newpage
\bibliography{WDBD_ZTF0038_clean}{}
\bibliographystyle{aasjournalv7}

\appendix

\onecolumngrid


 \section{Eclipse Fit Corner Plot}\label{sec:eclipse_profile_appendix}

Posterior distributions for the eclipse profile model parameters. The fitting procedure is discussed in Section \ref{subsec:eclipse}.
 
\begin{figure*}[h]
    \centering
    \includegraphics[width=1\linewidth]{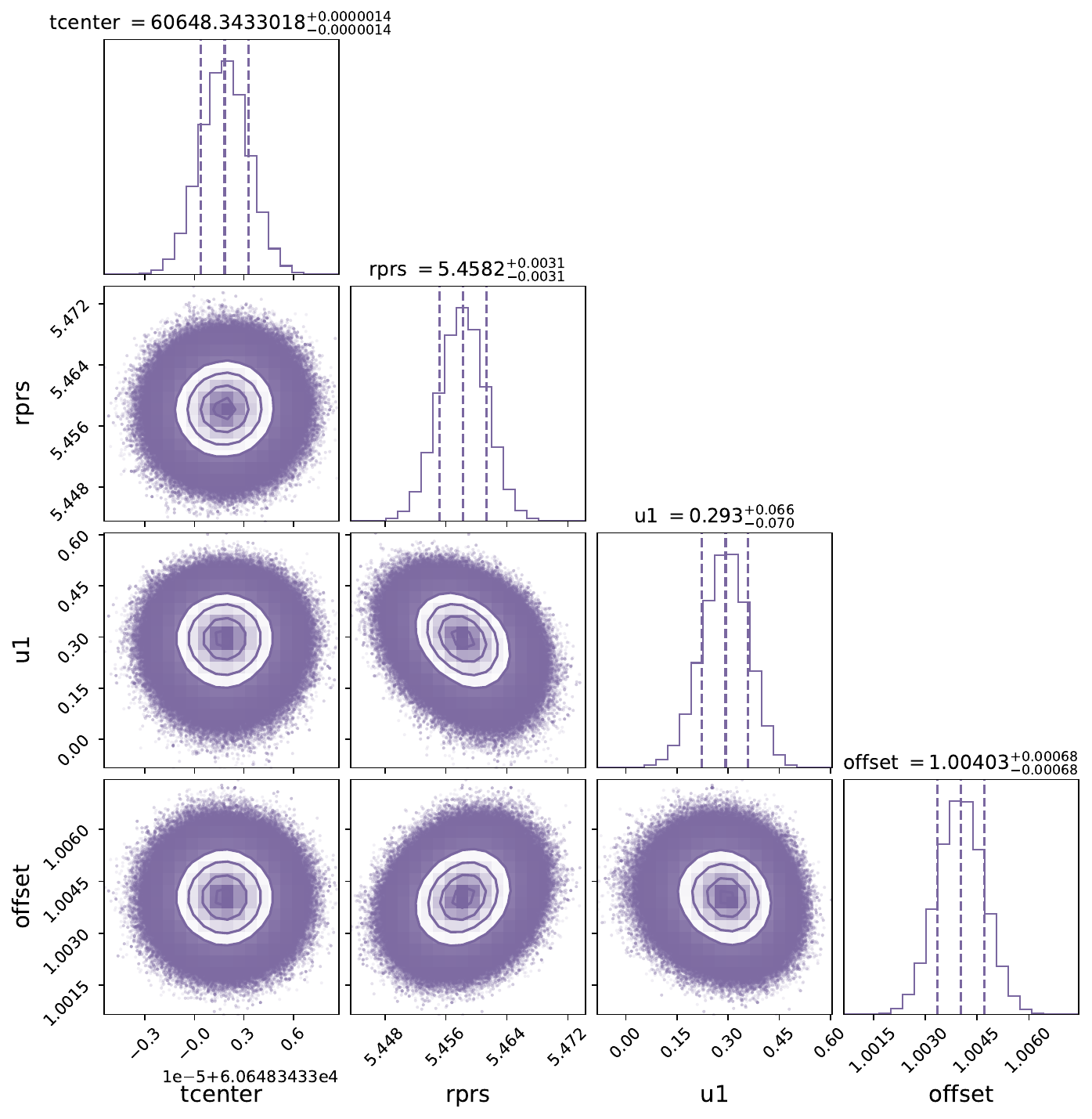}
    \caption{Corner plot of the MCMC eclipse fitting using a \texttt{batman} transit model. The best-fit model is shown in Figure \ref{fig:eclipse_fit}.}
    \label{fig:eclipse_fit_corner_plot}
\end{figure*}

\onecolumngrid
\clearpage
\section{Sonora Elf Owl Corner Plot}

Posterior distributions from fitting the nightside spectrum of ZTF0038B  with the Sonora Elf Owl model grid \citep{Mukherjee2024}, as discussed in Section \ref{subsec:forward_modeling}.

\begin{figure*}[h]
    \centering
    \includegraphics[width=1\linewidth]{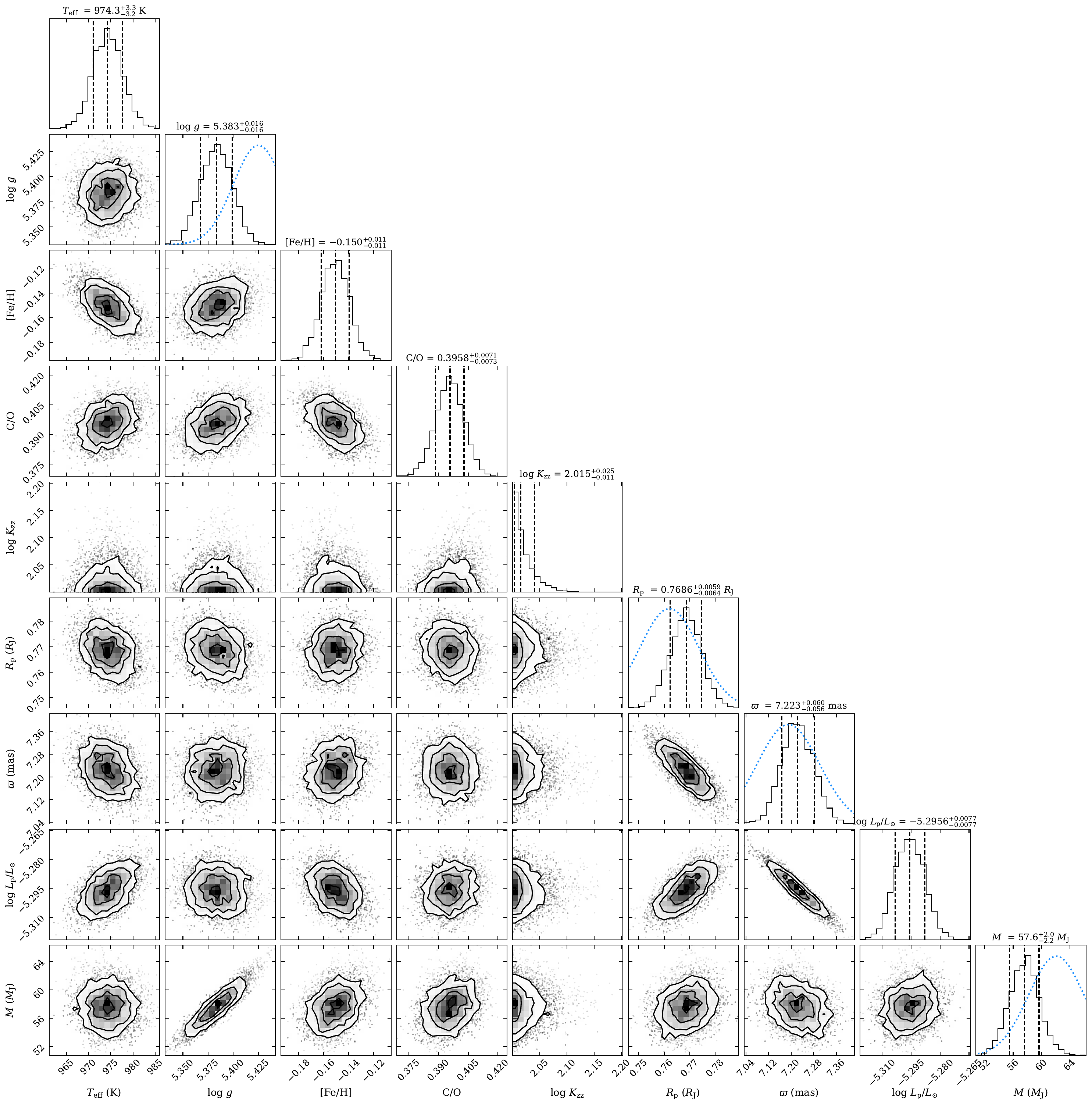}
    \caption{Corner plot showing the posterior distributions from nested sampling of the Sonora Elf Owl grid fit to the brown dwarf's nightside spectrum \citep{Mukherjee2024}. We use normal priors for log($g$), R$_p$, $\omega$, and $M$, as shown blue dashed lines in the relevant parameter histograms. }
    \label{fig:elf_owl_corner_plot}
\end{figure*}

\end{document}